\documentclass[12pt,preprint]{aastex}

\usepackage[usenames]{color}

\usepackage{graphicx}
\usepackage{epstopdf}
\usepackage{url}

\newcommand{\be}{\begin{equation}}
\newcommand{\ee}{\end{equation}}
\newcommand{\ba}{\begin{eqnarray}}
\newcommand{\ea}{\end{eqnarray}}
\newcommand{\bc}{\begin{center}}
\newcommand{\ec}{\end{center}}
\newcommand{\lsi}{LS~I~+61$^{\circ}$303}
\newcommand{\ls}{LS 5039}

\begin{document}

\shorttitle{\textsc{Long-term X-ray monitoring of \lsi}}
\shortauthors{\textsc{Li et al.}}

\title{\textsc{Long-term X-ray monitoring of \lsi: \\ analysis of spectral variability and flares}}

\author{Jian Li\altaffilmark{1}, Diego F. Torres\altaffilmark{2,3}, Shu Zhang\altaffilmark{1}, Yupeng Chen\altaffilmark{1}, Daniela Hadasch\altaffilmark{3},  Paul S. Ray\altaffilmark{4}, Peter Kretschmar\altaffilmark{5}, Nanda Rea\altaffilmark{3}, \& Jianmin Wang\altaffilmark{1,6}
}

\altaffiltext{1}{Laboratory for Particle Astrophysics, Institute of High
Energy Physics, Beijing 100049, China. Email: jianli@ihep.ac.cn}
\altaffiltext{2}{Instituci\'o Catalana de Recerca i Estudis Avan\c{c}ats (ICREA).}
\altaffiltext{3}{Institut de Ci\`encies de l'Espai (IEEC-CSIC),
              Campus UAB,  Torre C5, 2a planta,
              08193 Barcelona, Spain}
\altaffiltext{4}{Space Science Division, Naval Research Laboratory, Washington, DC 20375-5352}
\altaffiltext{5}{ESA-European Space Astronomy Centre, 28691 Villanueva de la Ca\~nada, Madrid, Spain}
\altaffiltext{6}{Theoretical Physics Center for Science Facilities (TPCSF), CAS, China}
\begin{abstract}

We report on the full analysis of a \textit{Rossi X-ray Timing Explorer} (\textit{RXTE})
Proportional Counter Array (PCA) monitoring of the $\gamma$-ray binary
system \lsi. The data set covers 42 contiguous cycles of the
system orbital motion. Analyzing this X-ray monitoring dataset, the
largest to date for this source, we report on the variability of the
orbital profile and the spectral distribution, and provide strong
evidence for an anti-correlation between flux and spectral index (the
higher the flux, the harder the spectral index). Furthermore, we
present the analysis of two newly discovered ks-timescale flares,
which present significant variability also on shorter timescales, and
tend to occur at orbital phases between 0.6-0.9. However, a detailed
timing analysis of the flares does not show any coherent or
quasi-coherent (QPO) structure in their power spectra. We also
investigated the possible appearance of the radio super-orbital
modulation at X-rays energies, but we could not unambiguously detect
such modulation in the system flux history, nor in the evolution of
its orbital modulation fraction.

\end{abstract}


\section{Introduction}

Among the class of X-ray binaries, \lsi\  is peculiar  because it was detected in all energy ranges from radio to TeV.
The other binary systems with somewhat similar characteristics are PSR B\,1259-63 (Aharonian et al. 2005b), a system hosting a pulsar, and Cyg X--1 (Albert et al. 2007), a system most likely hosting a black hole, and \ls\ (Aharonian et al. 2005a, Aharonian et al. 2006).
The nature of the compact object in both \ls\ and \lsi\ remains unknown.
We do know that \lsi\ consists of a rapidly rotating B0 Ve star with an equatorial outflowing disk orbited by a low mass ({\it M}$\sim$1--4\,$M_{\sun}$) compact object with
a period of $\sim$26.5 days in an eccentric orbit
(see Casares et al. 2005; Grundstrom et al. 2007; Aragona et al. 2009).  This periodicity is
visible in the optical band (Hutchings \& Crampton 1981, Mendelson \& Mazeh 1989), in the
infrared (Paredes et al. 1994), in soft X-rays (Paredes et al. 1997), in the
$H_{\alpha}$ emission line (Zamanov et al. 1999), and in the high-energy band
(20\,MeV--100\,GeV; Abdo et al. 2009).
The radio outbursts not only show this 26.5 day periodicity but also a shift of the phase of the
radio maximum between 0.45$\leq$$\Phi$$\leq$0.9, with a 1667$\pm$8 day second modulation (Gregory 2002).
In the X-ray band, a number of
observational campaigns have been carried out, showing that the emission is variable and has a non-thermal
spectrum. Two simultaneous X-ray and radio observations
showed that the X-ray
emission peaks almost half an orbit before the radio and, in both bands, the
orbital phases of the flux maxima seemingly drift from orbit-to-orbit (Taylor et al. 1996; Harrison 2000). However, soft X-ray pointed observations of \lsi\ have in general been too limited to cover full orbits of the system or to study long-term evolution of the X-ray orbital profile. This limitation was recently removed 
by the work of Torres et al. (2010, referred here as Paper I) who showed
that the periodic behavior is visible only
over long integration times and that profile variability is seen from orbit-to-orbit all the way up to multi-year timescales, with the phase of the profile maximum also changing.

To explain the behavior and
constituency of \lsi\, there are two major models: 1) Accretion onto a compact object, a black hole or a neutron star (Taylor \& Gregory 1984), and 2) Interaction of a young rotation powered pulsar with the wind from the companion Be star (Maraschi \& Treves 1981). Both Sidoli et al. (2006) and Rea et al. (2010) searched for evidence of pulsations (which in principle could support
either model) but found none. To prove the accretion model, one should
make direct measurements of accretion signatures or of jet like
structures. In 2004, Massi et al. (2004) reported on the existence of a large-scale
persistent radio jet. But Albert et al. (2008) did not find any
such jet at any of the scales explored using radio interferometric
observations from the MERLIN, EVN, and VLBA arrays in 2006. In radio
images of \lsi\ obtained during a complete orbital cycle, Dhawan et al. (2006) saw an extended
feature, which they ascribed to a cometary tail that changes orientation along the orbit. The same structure was confirmed several months later by Albert et al. (2008). But neither of these imaging
studies found a clear evidence for a jet.
Another proof for the accretion/jet models would be the discovery of a cutoff in the
10--100\,keV energy spectrum.
But studies made by Chernyakova et al. (2006) with {\it
{\it XMM}-Newton} and {\it INTEGRAL} data could not bring this proof.
Recent investigations on a much larger  {\it INTEGRAL} dataset reached the same conclusion (Zhang et al. 2010).
The data
could be fitted well by a simple power law without features,
matching to the higher energy spectrum in the 100\,keV--10\,MeV band found with
OSSE (Tavani et al. 1996) and COMPTEL (van Dijk et al. 1996). If the system is an accreting neutron star or black hole, one expects to find a cut-off power-law spectrum in the hard X-ray band with a cut-off energy normally at 10--60 keV for neutron stars (e.g. Filippova et al. 2005) and at $\sim$ 100 keV for black holes (McClintock \& Remillard 2003). Even when one can think of situations where the accretion disc is masked by a jet component, producing a featureless power law, if the accretion flow and the jet would give comparable contributions in the hard X-ray band, one would still expect at least a feature in the INTEGRAL energy range, due to the disappearance of contribution of the disc at high energies (see e.g., Chernyakova et al. 2006, Zhang et al. 2010).
Sidoli et al. (2006) could not find proofs for accretion either, although they found (in
{\it XMM}-Newton and BeppoSAX observations) evidence for a rapid change in flux in
timescales of hours. This behavior has also been observed in {\it Chandra} data (Paredes et al. 2007, Rea et al. 2010).
Esposito et al. (2007) carried out  a {\it Swift}/XRT monitoring,
observing flux fluctuations up to a factor of 3 at intra-hour scales, but once
again, finding no emission or absorption lines in the spectrum.
An {\it RXTE}  long-term monitoring, from August 28 2007 to February 2 2008 (a dataset which is included in our analysis)
was presented by Smith et al. (2009). They hinted at
a correlation between the flux and photon index, with the spectrum
becoming harder at higher fluxes.
Furthermore they hinted at the orbital modulation, peaking at phases 0.6--0.7 as claimed by Paredes et al. (1997)
and Esposito et al. (2007), which was seemingly not smooth, but presented short-timescale flares and
very strong orbit-to-orbit variability. The short time intervals covered in these publications did not permit strict conclusions, some of which were later to be claimed in Paper I and are studied in more detail here.

Finally, we remark that \lsi\ has also been extensively studied  in the very high energy range (VHE). The
MAGIC collaboration first detected \lsi\ as a variable TeV source, which was
independently confirmed by the VERITAS collaboration (Albert J. et al. 2006; Acciari et al. 2008). Both experiments found TeV emission only near the apastron
passage of the compact object in its orbit around the Be star.  The MAGIC
collaboration conducted a multiwavelength campaign including the MAGIC
telescope, {\it XMM-Newton} and {\it Swift} during 60\% of an orbit in 2007
September (Anderhub et al. 2009).
They discovered an X-ray/VHE gamma-ray correlation during an outburst around orbital phase 0.62, while the uncertainties prevent to be sure about the existence of the correlation outside the outburst.
Again, this conclusion was also derived by the VERITAS collaboration, when reporting the lack of X-ray/VHE correlation based on contemporaneous data with a VHE sampling that is not dense enough (Acciari et al. 2009).
\lsi\ even seems to have disappeared in the VHE band during quite some time,
following recent results presented by VERITAS  (Aliu et al. 2010, see below).

Long-term and especially continuous monitoring of gamma-ray binaries has been a goal for many years, so that we can start to see differences between them that could help us distinguish their nature, the possible distinct influence of the companion star, and the possible variability expected at other frequencies where only snapshots are possible. Long-term monitoring provides information on possible trends in the overall behavior of the source, and gives perspective as of the steadiness or variability of the former conclusion. Such monitoring can also catch unusual events or source states, or duty cycles, that could provide the key to the nature of the source. There are a few potential observations that one may think would conclusively demonstrate the true nature, such as the discovery of pulsations at any frequency or the detection of clear accretion signals like accretion lines, both of which are for the moment still elusive (see, e.g., Rea et al. 2010).
Multifrequency observations can provide knowledge of the dominance of single or multiple
particle populations, and of the nature of these particles. Thus, several questions can be tackled with this campaign. Some examples are whether the hinted anti-correlation between spectral index and flux is real, whether it changes with time if so, whether the spectra deviates from a pure absorbed power law along the different observation timescales, whether the anti-correlation between flux and spectral index is also present in flares, whether the modulation fraction varies, whether there were additional flares from the source,
whether they come at particular phases of the orbit and which is their inner structure, whether they had timing signatures, and others. We tackle these questions here.

This paper is thus a detailed follow up of Paper I, presenting the full analysis of the largest dataset on \lsi\ obtained in soft X-rays to date, enlarging that used in Paper I by several orbits, and
concentrating  on the spectral behavior of the source. Analyzing this  X-ray monitoring dataset, we report on the variability of the orbital and spectral profiles,  and on an anti-correlation between flux and spectral index (the higher the flux, the harder the spectral index). Furthermore, we present timing analysis of two discovered flares, for which do not show any peculiar feature in their power spectra. These flares, given the PCA field of view, may or may not be associated with \lsi. However, we analyze all flares together (a total of 5 are known coming from the general direction of the system) and find circumstantial evidence for an association.

\section{Observations and data analysis}

Our dataset covers the period between 2006 October and 2010 September, containing and enlarging the data used for Paper I  and  it includes 378
{\it RXTE}/PCA pointed observations
identified by proposal numbers 92418, 93017, 93100, 93101, 93102, 94102 and 95102. As in Paper I, here we
focus on the data starting on 2007 September because of the smaller time gaps between observations, providing a total exposure time of 549 ks on the source.

In the analysis of PCA data two separate data modes were used: the ``Standard 2'' mode was used for lightcurve and spectral analysis with 16s accumulation time and 129 energy channels; for analysis of flares, which required greater time resolution, the ``Good Xenon'' mode was utilized with 1 $\mu s$ resolution, and 256 energy channels.
Data reduction in both modes was performed using HEASoft 6.6. For the Standard 2 analysis we filtered the data using
the standard {\it RXTE}/PCA criteria. As explained by Smith et al. (2009), since combining large spectral data sets with differing PCA configurations (with associated differing calibrations) can produce large systematic errors, it is preferable to use a single common PCU for all available data sets.
PCU2 (in the 0--4 numbering scheme) is used for most of the
analysis as it was the only PCU that was 100\% on during the observations. We select time intervals where the source elevation is $>10^{\circ}$ and the pointing offset is $<0.02^{\circ}$.
PCA background lightcurves and spectra were generated using FTOOLS task {\tt pcabackest}. {\tt pcarsp} is used to generate PCA response matrices for spectra.
The background file used in analysis of PCA data is
the most recent available  from the HEASARC website for faint sources,
and detector breakdown events have been removed.\footnote{The background file is
\url{pca_bkgd_cmfaintl7_eMv20051128.mdl} and see the website:
\url{http://heasarc.gsfc.nasa.gov/docs/xte/recipes/pca_breakdown.html}
for more information on the breakdowns. The data have been barycentered using the {\tt
FTOOLS} routine {\tt faxbary} using the JPL DE405 solar system ephemeris.
}

As stated, we have analyzed the flares using Good Xenon data which are also barycentered using the FTOOLS routine {\tt faxbary} assuming in this process that they come from \lsi. We
caution that the PCA field of view is about 1 degree (FWHM)
thus providing no direct evidence for the relation of this flaring phenomenology with \lsi.
Since  none of these short flares were observed by any instrument with a better
spatial resolution,  we cannot exclude that they were generated
in a nearby (in sky projection) source.
During the observations of the two new flares -- at about MJD 54670.844444 and MJD 54699.653333 --
PCU 3 and 4 were on as well. To include more photons, lightcurves and spectra from top layers of PCU 2, 3, and 4 were extracted from 3--10 keV. Above 10 keV, there was a severe lack of photons leading
to large error bars in the energy spectrum that were not proper for
fitting.
Background subtraction for Good Xenon lightcurves was carried out using the FTOOLS task lcmath: it
subtracts a background lightcurve produced from Standard 2 data with the same combination of PCUs and the same energy range which was carried out by using FTOOLS task {\tt lcmath}.
Timing analysis of 0.01 seconds binned lightcurves was carried out using {\tt powspec}, producing power spectra in the range from 0.1 to 50 Hz.

\section{Results}

\subsection{Spectral analysis}

\begin{figure}[t]
\centering
  \includegraphics[angle=0, scale=0.85] {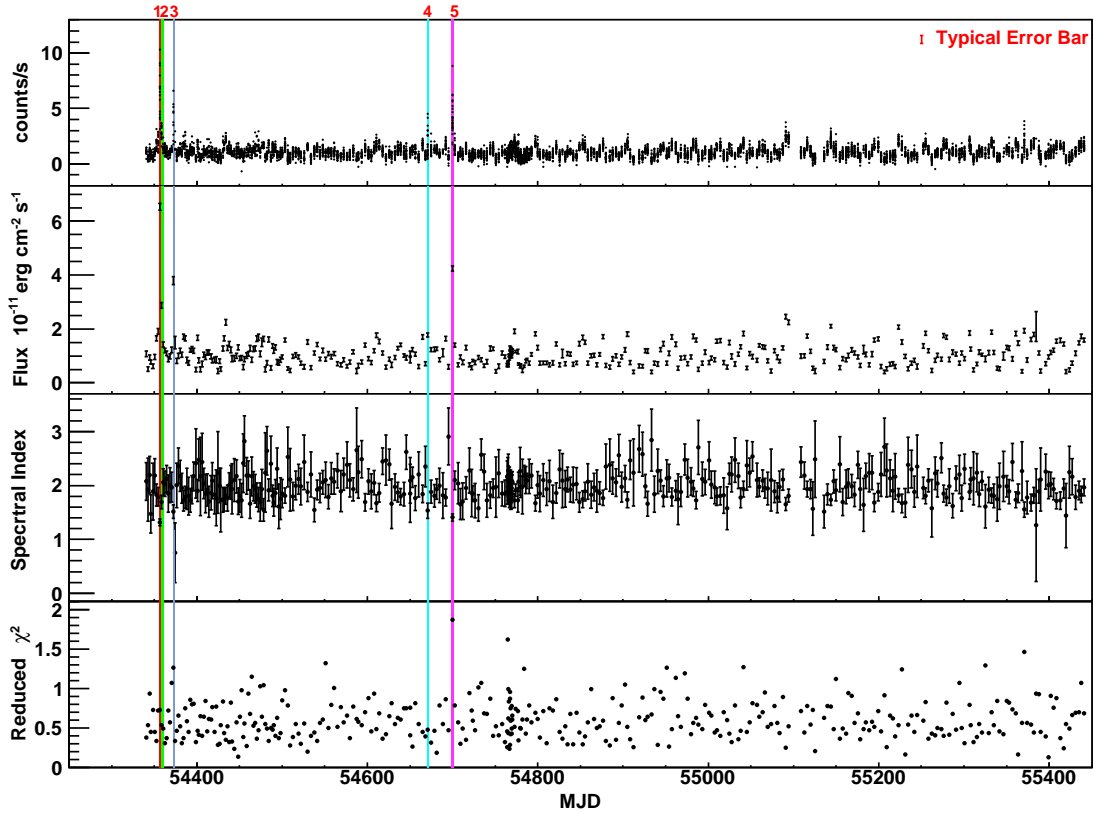}
     \caption{3--10 keV PCA count rate (first panel), observed flux in the 3-10 keV energy band (second panel), spectral index (third panel), and reduced $\chi^2$ (lower panel) from fitting each observational ID with a powerlaw function. Details are provided in the text. The vertical lines report on the flares periods; the last two are separately analyzed in the text. }
\label{f1}
\end{figure}

In the top panel of Figure \ref{f1}, we show the count rate in 64 second time bins in
the 3--10 keV band over the observation period.
For each observational ID, a simple power law shape, with absorbing hydrogen column density fixed at $0.75\times10^{22}$ cm$^{-2}$ (Kalberla et al. 2005, Smith et al. 2009), was used to fit the Standard 2 data. We adopt this most recent determination of $N_H$ and the spectral fit
used is
\begin{equation}
N(E) = K e^{-N_H\,\sigma(E)} \left( \frac{E}{{\rm keV}}\right)^{-\alpha},
\end{equation}
where $K$ is a normalization at 1 keV, $\sigma$ is the photoelectric cross section and $\alpha$ is the photon index, referred to as slope. Higher values of alpha are
referred to as being \emph{soft} and lower values as being \emph{hard}.
The flux values were obtained by integrating the best fit spectra and 1$\sigma$ error levels were generated.
In Figure \ref{f1}, we show this derived flux and the photon index in the second and third
panels. The reduced $\chi^2$ values for the fit are also shown in Figure \ref{f1}: there are 368 data points averaging around 0.60$\pm$0.01. All errors in this paper are reported in a significance of 1 $\sigma$.

As previously seen in Paper I, the lightcurve shows two new flares in addition of the three events previously reported  by Smith et al.(2009).  The average flux value in energy units, plotted in the second panel of Figure \ref{f1}, is {(1.024$\pm$0.004)}$\times$10$^{-11}$erg cm$^{-2}$s$^{-1}$. Whereas fitting a constant to the lightcurve shows obvious variability of the flux (the reduced $\chi^2$ is {1.3$\times 10^4$/367)}, the variability in the spectral index appears only moderate. This can be seen in the third panel of Figure \ref{f1}:
the reduced $\chi^2$ of a horizontal line is {426/367} and
corresponds to the spectral index being a non-zero constant value
during the observational period with a significance of $\sim2.4 \sigma$.
The spectral index averaged over the all {\it RXTE}/PCA observations is {1.876$\pm$0.012.}

\begin{figure}[t]
\centering
   \includegraphics[angle=0, scale=0.75] {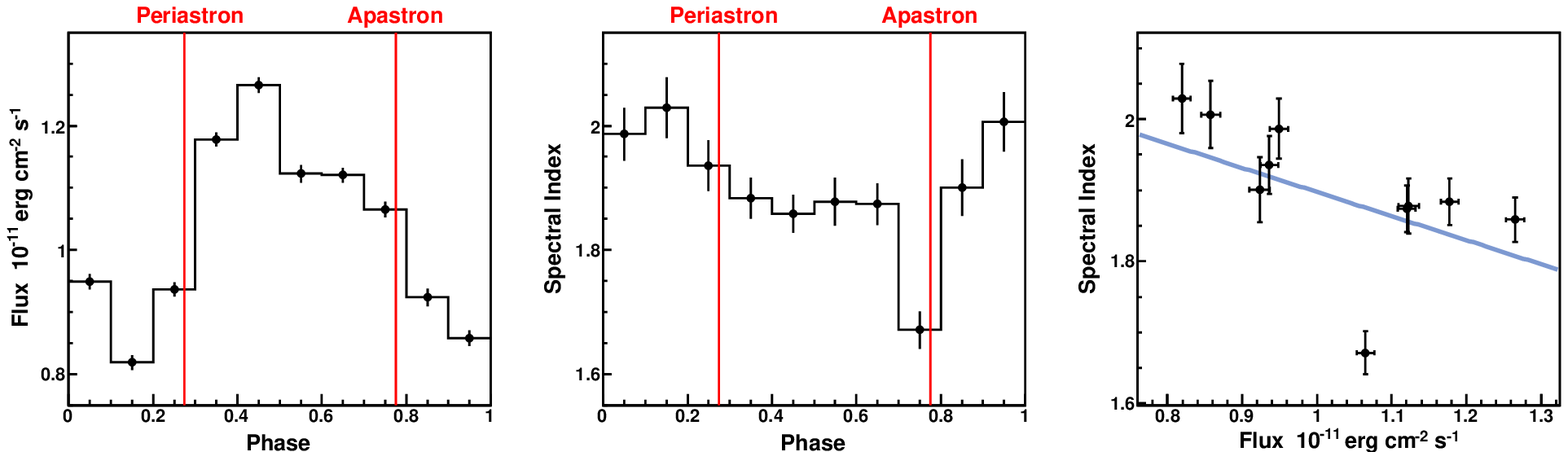}
 \includegraphics[angle=0, scale=0.75] {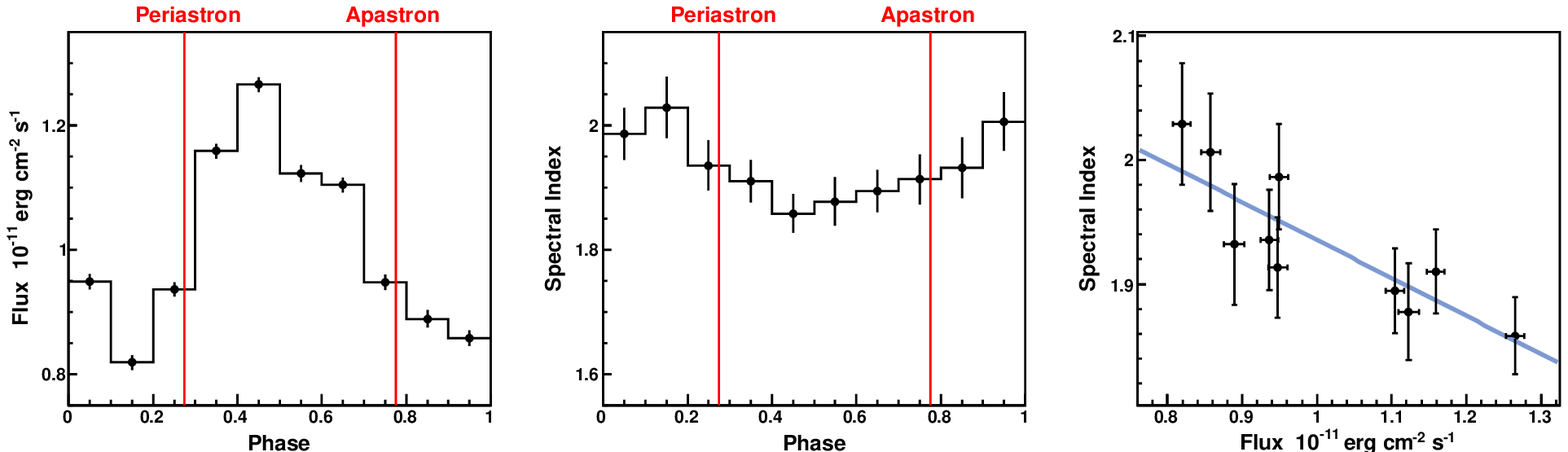}
     \caption{Top: The folded orbital profile for the 3--10 keV flux (left) and spectral index (middle) derived for all {\it RXTE} data (including the flares periods). The rightmost panel represents the relation between the values of the spectral index and the flux determined for each of the
10 phase bins in which the orbit was
divided. The line is a linear fit to the data as described in the text. Bottom: Same as the top panels but excluding the 5 flares from the data.}
\label{f2}
\end{figure}

In order to produce orbital profiles, we folded the data by the measured radio period of 26.4960$\pm$0.0028 days (Gregory 2002) and an orbital zero phase at MJD 43366.275 (Gregory \& Taylor 1978) and then divided the data into 10
orbital phase bins. Figure \ref{f2} shows these orbital profiles for the data
both with (top row) and without (bottom row) the 5 flares. The left
panels show the flux vs. phase, the middle show spectral index vs. phase,
and the right panels show spectral index vs flux for the 10 phase bins. A clear orbital modulation can be seen in the flux data (see left panels of Figure \ref{f2}), with a peak rising  up around phase 0.45. This phase is shortly after the periastron passage, which is between phase 0.23--0.30 (Casares et al. 2005; Aragona et al. 2009; Grundstrom et al. 2007). Note in the top middle panel of Figure \ref{f2} that there is a very hard value of the spectral
index in the phase bin 0.7--0.8. This dip is due to two flares with a very
hard spectral slope that reduce the average value of the slope in this
bin. In the right panel of Figure \ref{f2}, we show a clear anti-correlation
between flux and spectral index, i.e., the higher the flux, the harder
the spectral index. Smith et al. (2009) also hinted at an anti-correlation between the flux and the spectral index but using only {\it RXTE} data covering the 2007-2008
campaign (MJD 54340-54500). This entire campaign is included in our
study and we explore this anti-correlation with this much larger dataset.
In the rightmost panel of Figure \ref{f2}, we show a linear fit to the data.
In the top right panel (flares included), the linear fit has a reduced $\chi^2$ of {57.4/8} 
and a slope of {$-0.34\pm0.1$} 
 while in the bottom right panel (flares excluded), the linear fit has
a reduced $\chi^2$ of {4.2/8} and a slope of {$-0.31\pm0.1$.}

Figure 3 explores this flux-spectral index anticorrelation
for the entire data set. The left panel shows all data, the
middle panel excludes the 5 flares, and the right panel shows data for
the 5 flares only.
From left to right, the linear fits to the
data have a slope of {$-0.14\pm0.01$}, with a reduced $\chi^2$ of {221.3/366};
{$-0.26\pm0.03$}, with a reduced $\chi^2$ of {192.9/361}; and
 $-0.06\pm0.03$, with a reduced $\chi^2$ of 2.4/3. 

Note that the two observation IDs with very low value of spectral
index and larger error bars, 93100-01-18-00 and 95102-01-54-00, are located at MJD 54374.15621064 and MJD 55384.41457248. In these observations, the good time intervals are only 160 s and 32 s respectively, quite smaller than the typical 1000 -- 2000 s of good time intervals obtained for the other data points. As a result, few photons  are accumulated in these short intervals, so the data points in the
energy spectrum have larger error bars and we cannot derive a good and reliable
fit, even when the $\chi^2$-value is small due to the errors. As a result of the unreliable fit in these two observations, we exclude these data from our analysis below.
To be sure that the found anti-correlation is not spurious, we
carefully examined two sources of bias: 1) systematic uncertainties in
the estimated background spectrum, and 2) extraction of the flux from the
fitted spectrum. Especially around 10 keV and
above, the background accounts for a large fraction of the observed counts. Thus, a small error in the background subtraction can have a large impact on the net counts of the source, which especially for
weak sources may lead to a significant change in the spectral index.
By using the {\tt corfile} and {\tt cornorm} commands in {\tt Xspec}, we estimated the largest uncertainty that may result from possible improper background subtraction and
found that the fitted fluxes and spectral indices, as well as their anticorrelation,
are not significantly affected. To check for effects of this second
possible bias, we extracted the lightcurves in count rate for each observation and found that the count rate versus the spectral index can be fitted by the same linear fits as
noted. 
The Pearson product-moment correlation coefficient\footnote{
See `Numerical recipes in C' available online at
http://www.fizyka.umk.pl/nrbook/bookcpdf.html
and especially,
Chapter 14.5 Linear Correlation for the coding tool, and Smith et al. 2009 or Albert et al. 2007b for applications.}
of all data including flare points
is \textbf{$-0.50_{-0.04}^{+0.04}$}, for data without flares it is
\textbf{$-0.50_{-0.04}^{+0.04}$}, and for flare points only it is
$-0.67_{-0.24}^{+0.57}$ (errors correspond to $1\sigma$ level). Taking into account the
non-linearity of errors, we find that the
correlation coefficent is different from zero by \textbf{$10.45\sigma$} 
 in the first case, by \textbf{$10.50\sigma$} 
 in the second (no flares) case, and by $1.1\sigma$ in the
third (only flares) case. Therefore, we again can conclude that there is a
significant anti-correlation between the spectral index and the flux using all data or just the data without the flares.
However, looking at the flares only, no anti-correlation can be claimed.

Comparing these three plots, one can conclude that the anti-correlation is an orbit-associated effect, with less presence in the flares. The data for the
flares show that an increase in flux by a factor of 3 can still be fitted by the same spectral index as the non-flare
data. This is not the case for the orbital-associated X-ray emission. This difference could indicate that either the process generating the flares is not the same from the usual X-ray emission of the source, or that the origin of the flares is not \lsi. We caution however that an alternative explanation to this, which is further explored below, is that the resulting behavior is a consequence of a single power law fit across the flare, when in reality, the spectral slope significantly changes within the bursts.

Using our calculated flux and
spectral index data, we explore whether the orbital profiles are stable
in time (in Paper I we did this for count rate only and in the wider range of
3-30 keV).
To do that exploration, we divided our data into periods of 6 months each, which provided good statistics
at all the phases of the orbit. The corresponding dates for
each 6-month period are given in Table 1.

Figure \ref{f4} shows the
evolution of the fitted flux (left panel) and the spectral index (middle
panel) during these periods from top to bottom, together with the evolution of the
relation between flux and spectral index (right panel), for the whole data
set excluding the flares. We have reconfirmed the significant variability in
the orbital profiles that was discussed in Paper I (which used only count
rates). Here we also see that there is some variability in the orbital
properties of the spectral index, although the significance of this
variability is less dramatic, in part hampered by the large error bars. As in Figure 1, the fits in the right panel explore the possible anti-correlation between flux and spectral index, which is again better found when the data without the flares is taken into consideration.

\begin{figure}[t]
\centering
  \includegraphics[angle=0, scale=0.28] {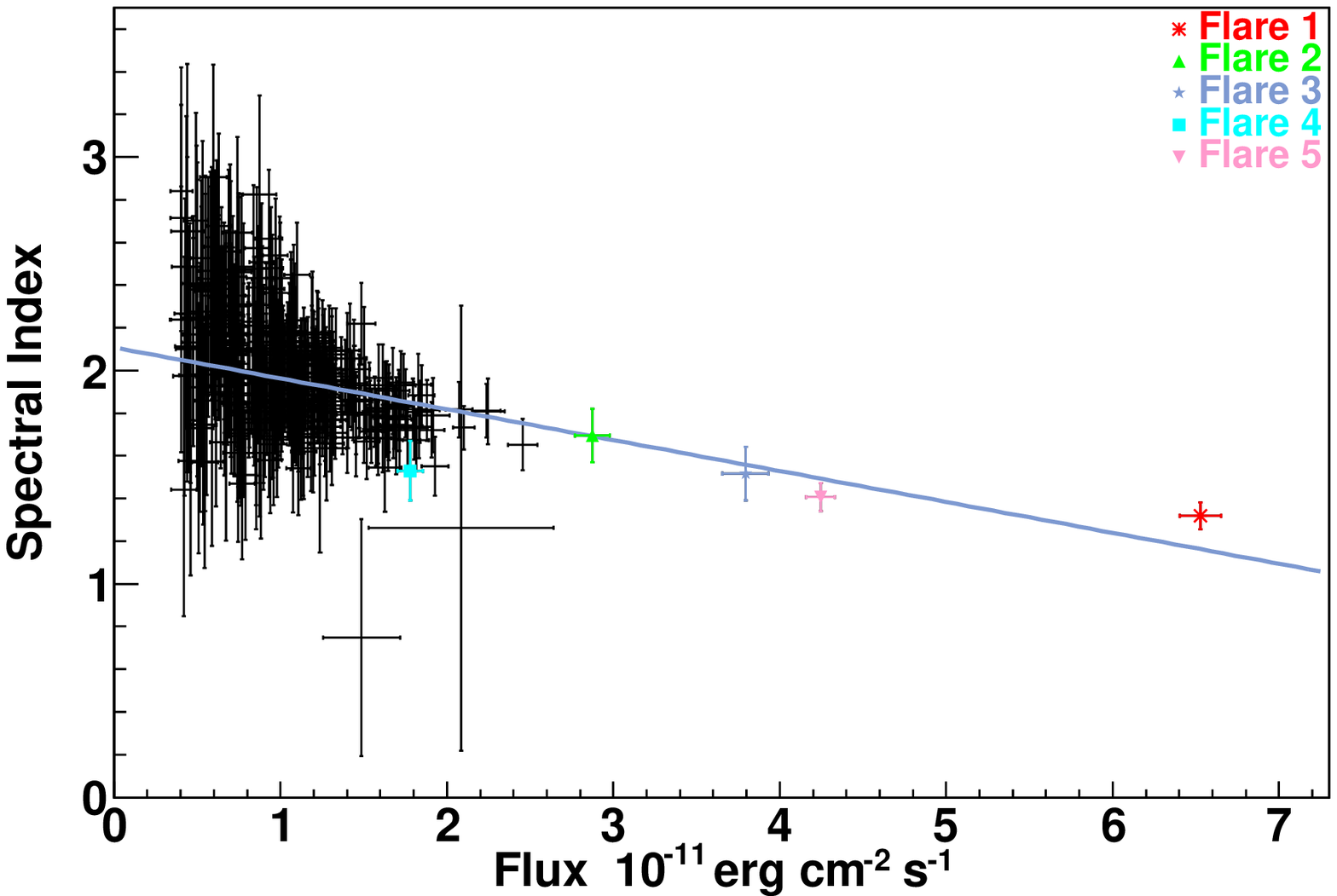}\hspace{-0.6cm}
    \includegraphics[angle=0, scale=0.28] {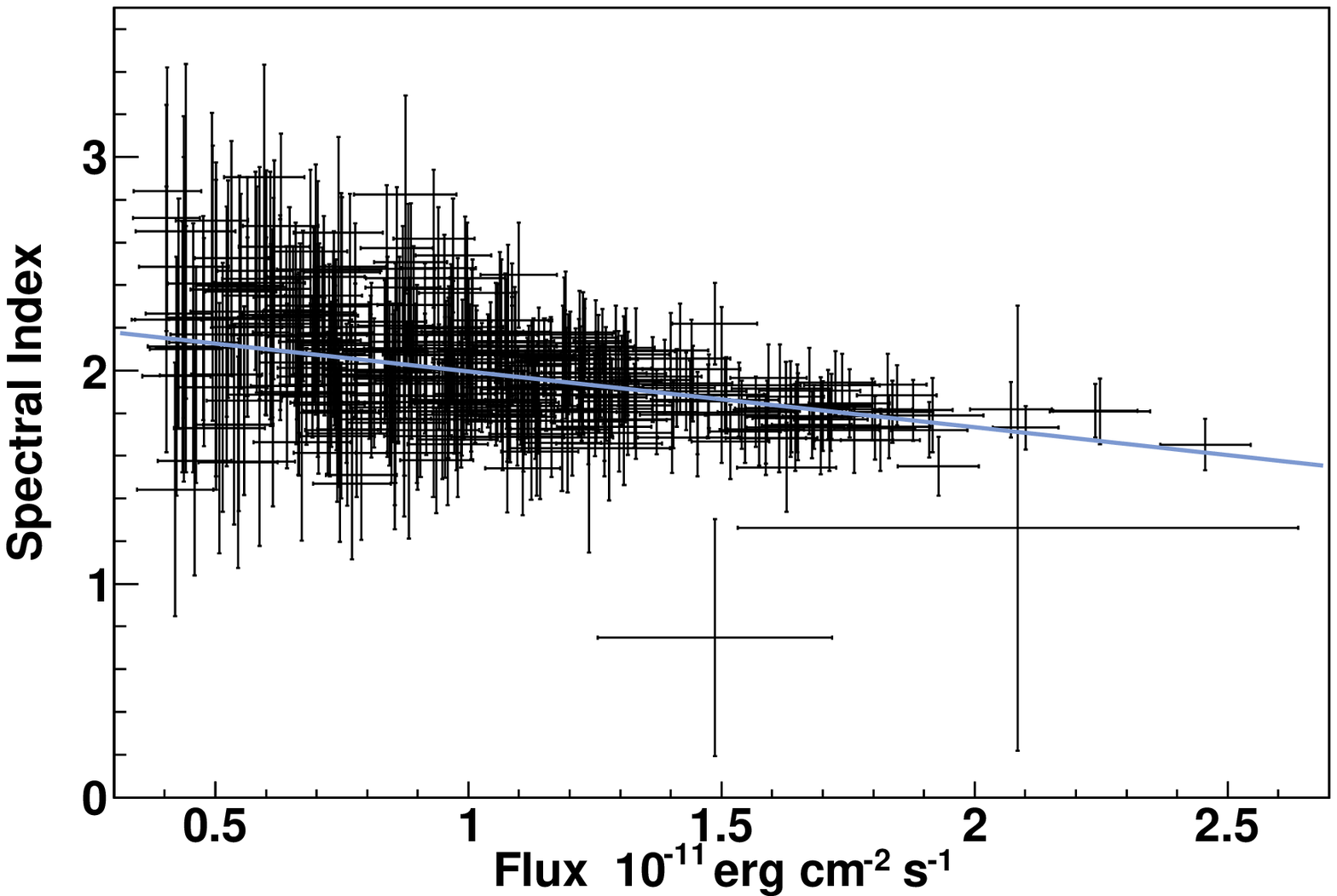}\hspace{-0.6cm}
      \includegraphics[angle=0, scale=0.28] {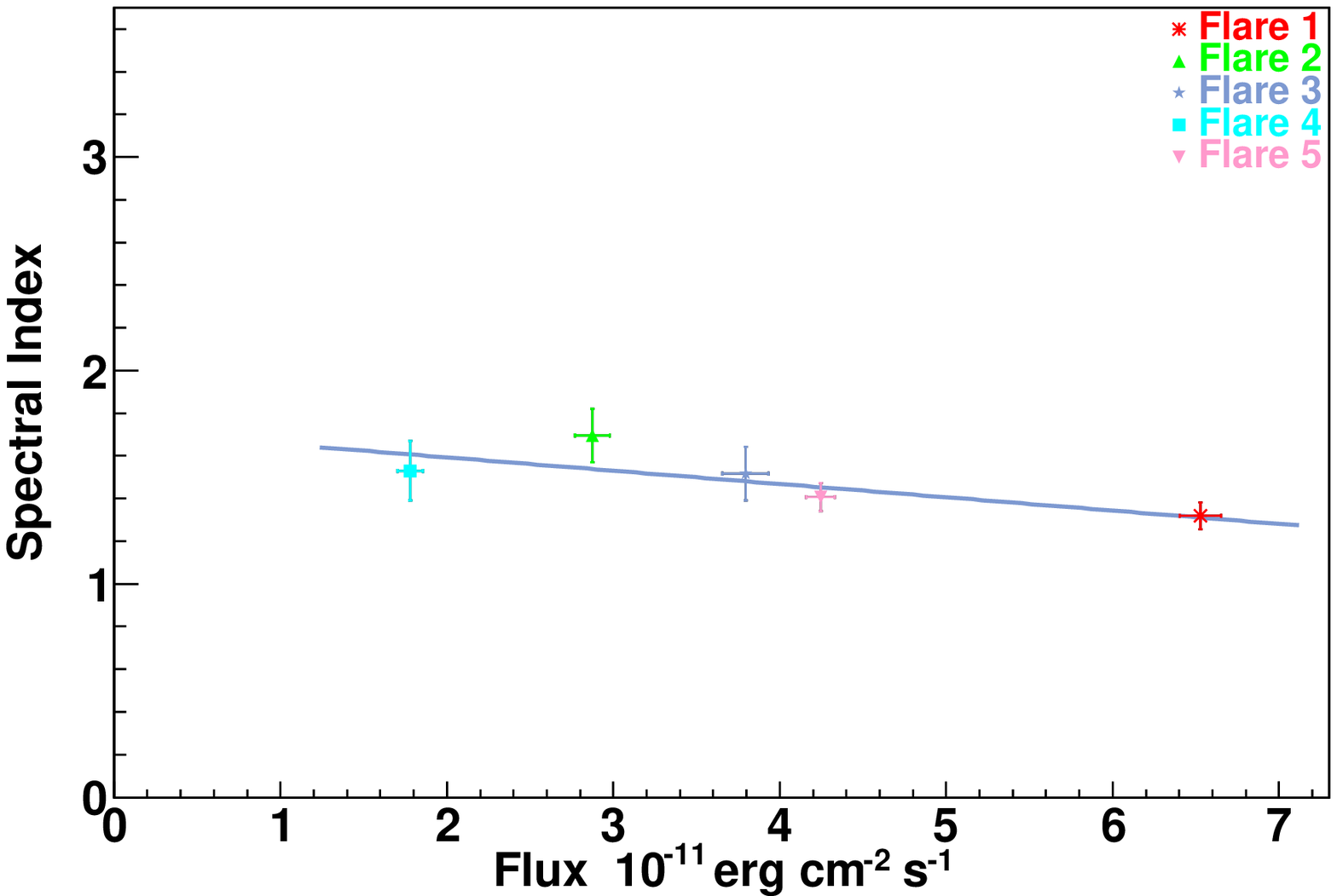}
     \caption{Spectral index vs. flux for the entire data set. The values for each data point are derived from fitting individual observational IDs of {\it RXTE} with a simple power law as explained in the text. The left panel shows the
whole data set, the middle panel shows the whole data set except for the
5 flares (which are marked in Figure 1), and the right panel shows the
5 flares only. The line in each panel is a linear fit to the data,
details of these fits are given in the text. }
\label{f3}
\end{figure}

\begin{figure}[t]
\centering
  \includegraphics[angle=0, scale=0.8] {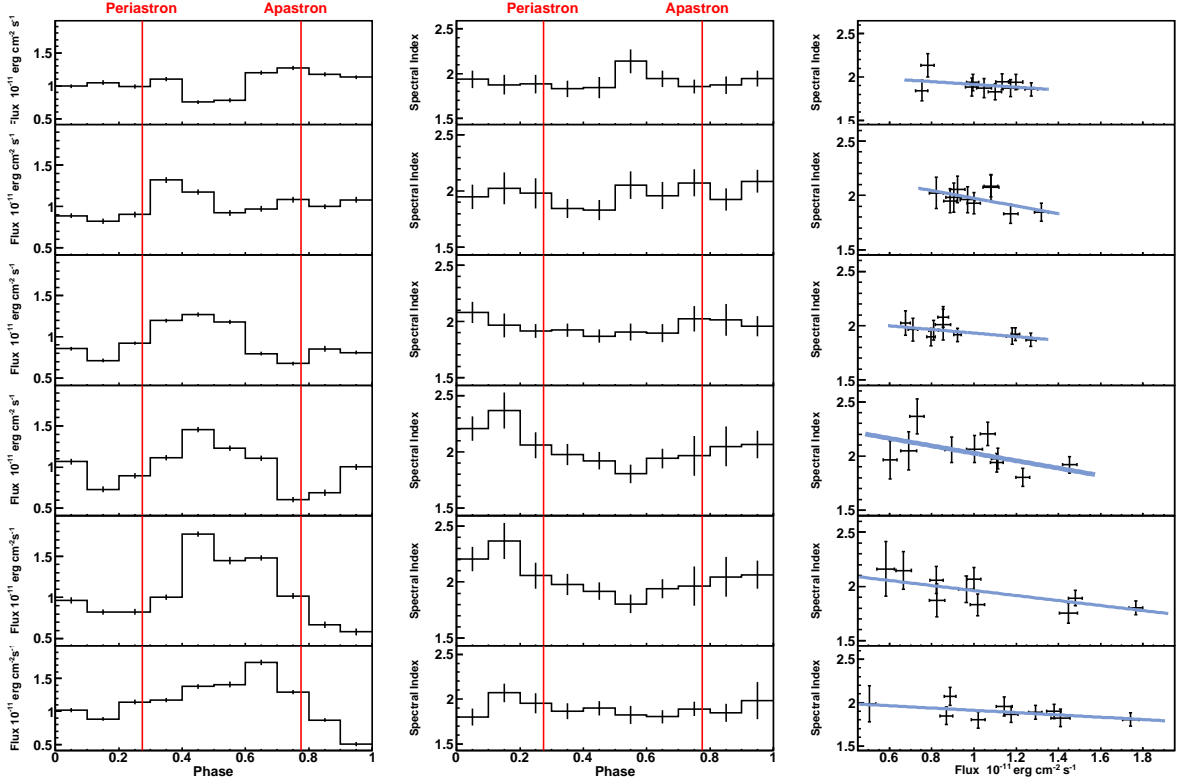}
 \caption{For each of the 5 separate 6-month periods considered, we show the 3-10 keV
lightcurves (left) and spectral indices (middle), excluding the
flares and folding at the orbital period. The right panel shows the spectral index versus flux relationship. }
\label{f4}
\end{figure}

In order to determine if this flux/spectral index anti-correlation is
maintained on all timescales, we also carried out an orbit-by-orbit
analysis for each of the 35 orbits covered by this data set. We found
that with the available data quality, the slope of the fit to the phase resolved
flux vs. the spectral index that is seen in Figure \ref{f3} is
essentially maintained. The average slope obtained for this orbit-by-orbit
analysis is \textbf{$-0.28 \pm 0.04$}. Thus we find that at all timescales,
from observation ID, to orbital, to months, and up to years, that as flux goes
up the spectral index goes down (i.e., it gets harder) and at a specific
slope.

\begin{table}[ptbptbptb]
\begin{center}
\label{tab1}
\caption{
For each 6-months time period, as
described in the text and shown in Figure \ref{f4}, we list the range of dates included, their corresponding MJDs,
the slope of the linear fit to the data (shown in Figure \ref{f4}, right panels), and the reduced
$\chi^2$ of the linear fit.
}
\vspace{5pt}
\small
\begin{tabular}{lllll}
\hline \hline
Time & Dates included & MJD &slope & reduced $\chi^2$\\
\hline
 1st 6-months &28/8/2007 --- 28/2/2008&54340-54524& -0.21$\pm$0.21&3.7/8\\
\hline
 2nd 6-months &29/2/2008  --- 29/8/2008&54525-54707& -0.35$\pm$0.22&4.8/8\\
\hline
3rd 6-months &30/8/2008  --- 01/3/2009&54708-54891& -0.17$\pm$0.12&3.2/8\\
\hline
4th 6-months &02/3/2009  --- 02/9/2009&54892-55076& -0.34$\pm$0.15 &11.2/8\\
\hline
5th 6-months &03/9/2009  --- 03/3/2010&55077-55258& -0.23$\pm$0.09 &5.7/8\\
\hline
6th 6-months &04/3/2010  --- 04/9/2010&55259-55443& -0.13$\pm$0.10 &4.6/8\\
\hline
\end{tabular}
\end{center}
\end{table}

\subsection{Flare Analysis}

Besides the three flares reported by Smith et al. (2009), we here report on the two additional ones
located at MJD 54670.844444 (observation ID 93102-01-25-01) and MJD 54699.653333 (observation ID 93102-01-29-01) that we have found in our data.
Following the numbering of Figure \ref{f1},  flares 1 through 5 happen at the \lsi\ phase
0.787, 0.861, 0.379, 0.651, and 0.739, respectively. While the statistics is not large, it
is interesting to note that 4 out of the 5 flares are grouped in
the 0.6--0.9 orbital phase bin of \lsi. In total over
the whole data set, there are 316
observations, and their distribution
into the 10 phase bins is as follows: we have 30,	31,	36,	37,	33,
26, 33, 33, 27 and 30 observations in each of the bins (from 0--0.1,
0.1--0.2, etc.). Using this distribution, we can
derive the probability of the observed configuration through the Binomial distribution. If the flares do
not correlate with the orbital phase, the probability of the observed
configuration is only $2.76 \times 10^{-3}$.
 The flare distribution in \lsi\ orbital phase bins might then point to a circumstantial evidence for an association, although statistics are still too
scarce to make a definite claim.

\begin{figure}[t]
\centering
  \includegraphics[angle=0, scale=0.38] {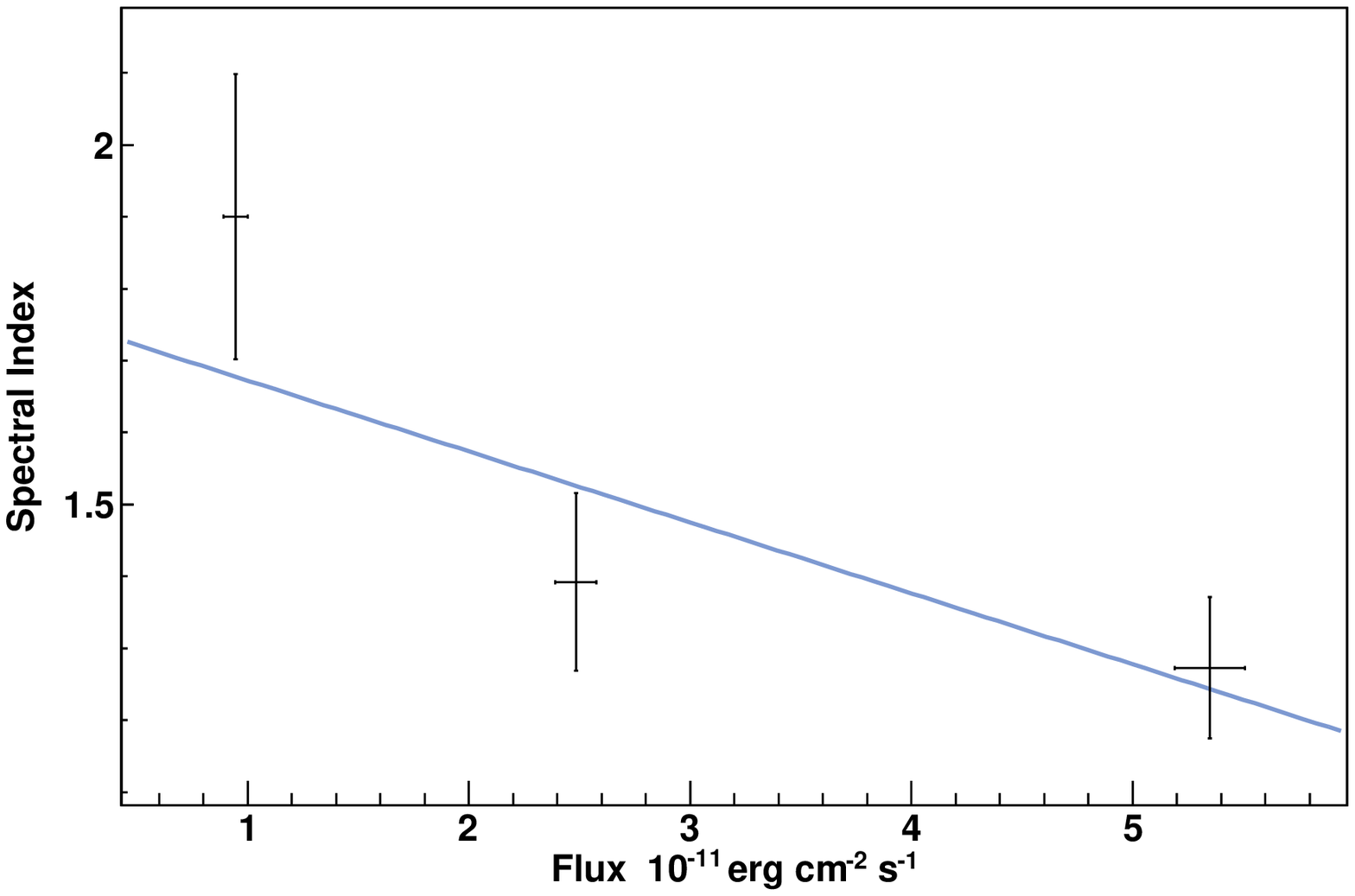}
    \includegraphics[angle=0, scale=0.38] {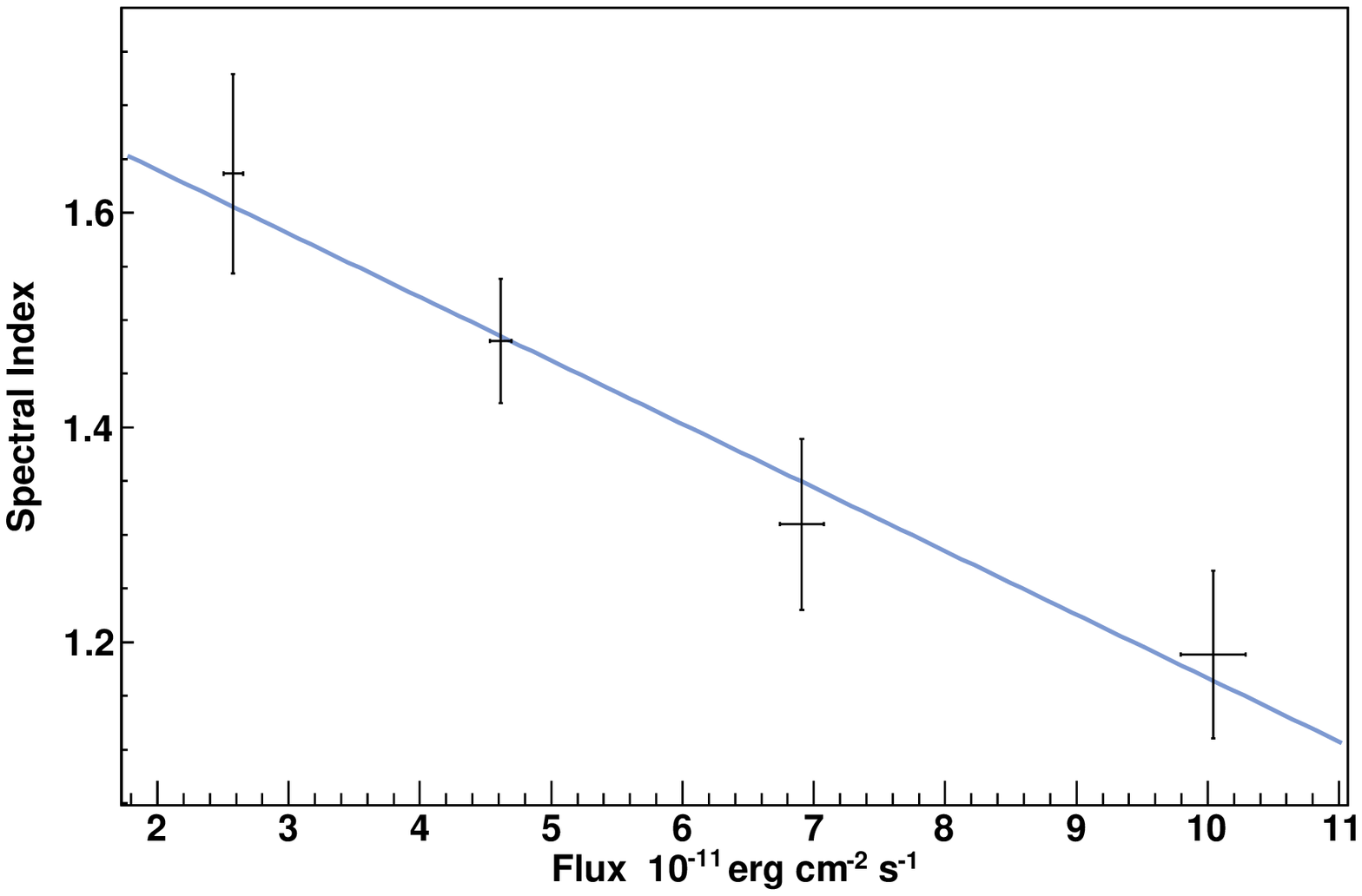}
     \caption{Spectral index versus flux for each of the two new flares analyzed, at MJD 54670.844444 and MJD 54699.653333, respectively. The linear fit is described in the text.}
     \label{f5}
\end{figure}

We also looked into the spectral details of the two new flares (for the others, see Smith et al. 2009) to
determine whether there is a correlation between spectral index and flux. The data on the two flares (4th \& 5th) were binned based on count rate levels.
The data for the 4th flare were binned into three groups with counts rates of 0--5, 5--12,  and $>12$ counts per second,
and the flux and spectral index were derived for these groups. The data for the 5th flare were binned for count rates in the ranges 0--10, 10--18, 18--26, and $>26$ counts per second.
For the two flares, the spectral index versus flux relation is shown in Figure \ref{f5}. Observe that the spectral index significantly
varies  during the flares, and that, despite the
the few degrees of freedom, a linear (anti) correlation is apparent. A linear fit of 4th flare produce a slope of $-0.1\pm0.04$ with a reduced $\chi^2$ of 2.52/1, while for 5th flare produced a slope of $-0.06\pm0.01 $ and a reduced $\chi^2$ of 0.46/2.

\begin{figure}[t]
\centering
    \includegraphics[angle=0, scale=0.8] {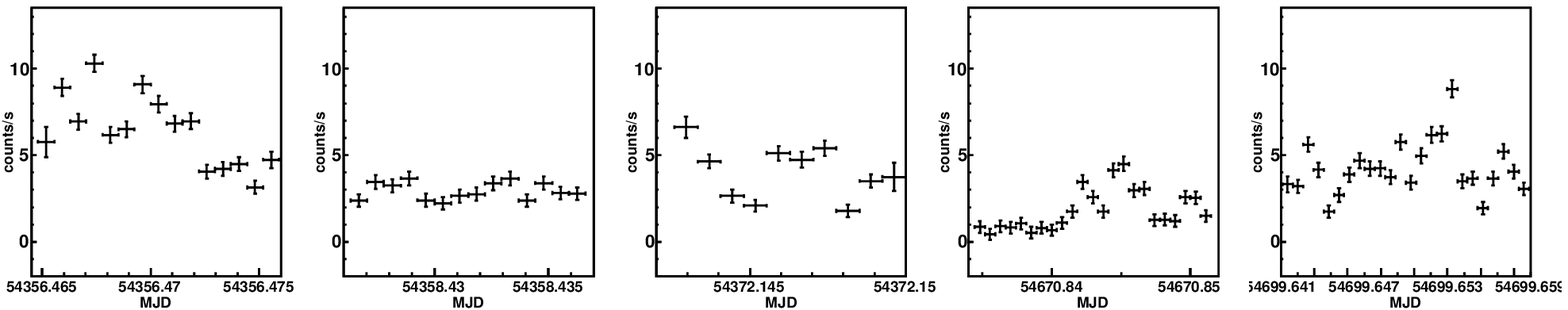}
     \includegraphics[angle=0, scale=0.8] {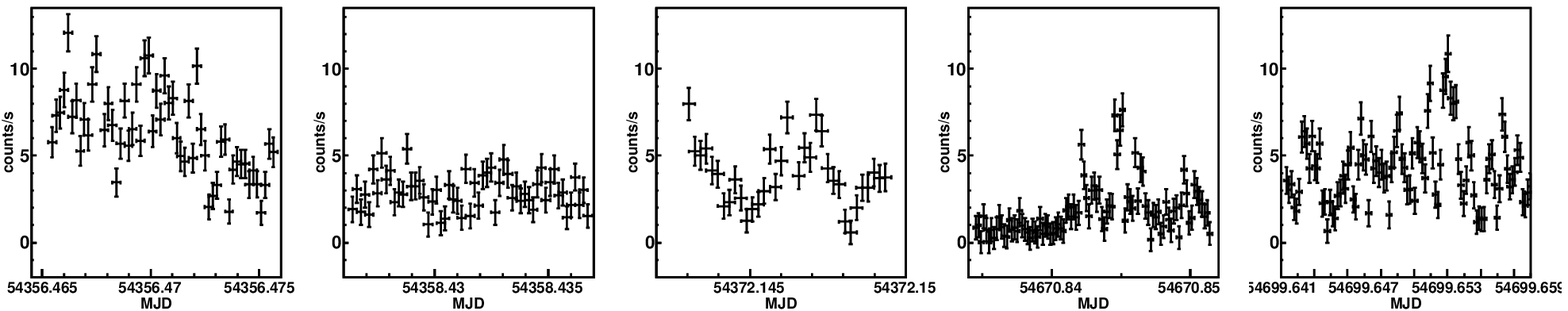}
     \caption{Zoom in the lightcurve presented in the top panel of Figure 1 for each of the 5 flares detected by PCA. Each point represents 64 s in the top panels and 16 s in the bottom panels.}
     \label{flare-counts}
\end{figure}

\begin{figure}[t]
\centering
  \includegraphics[angle=0, scale=0.38] {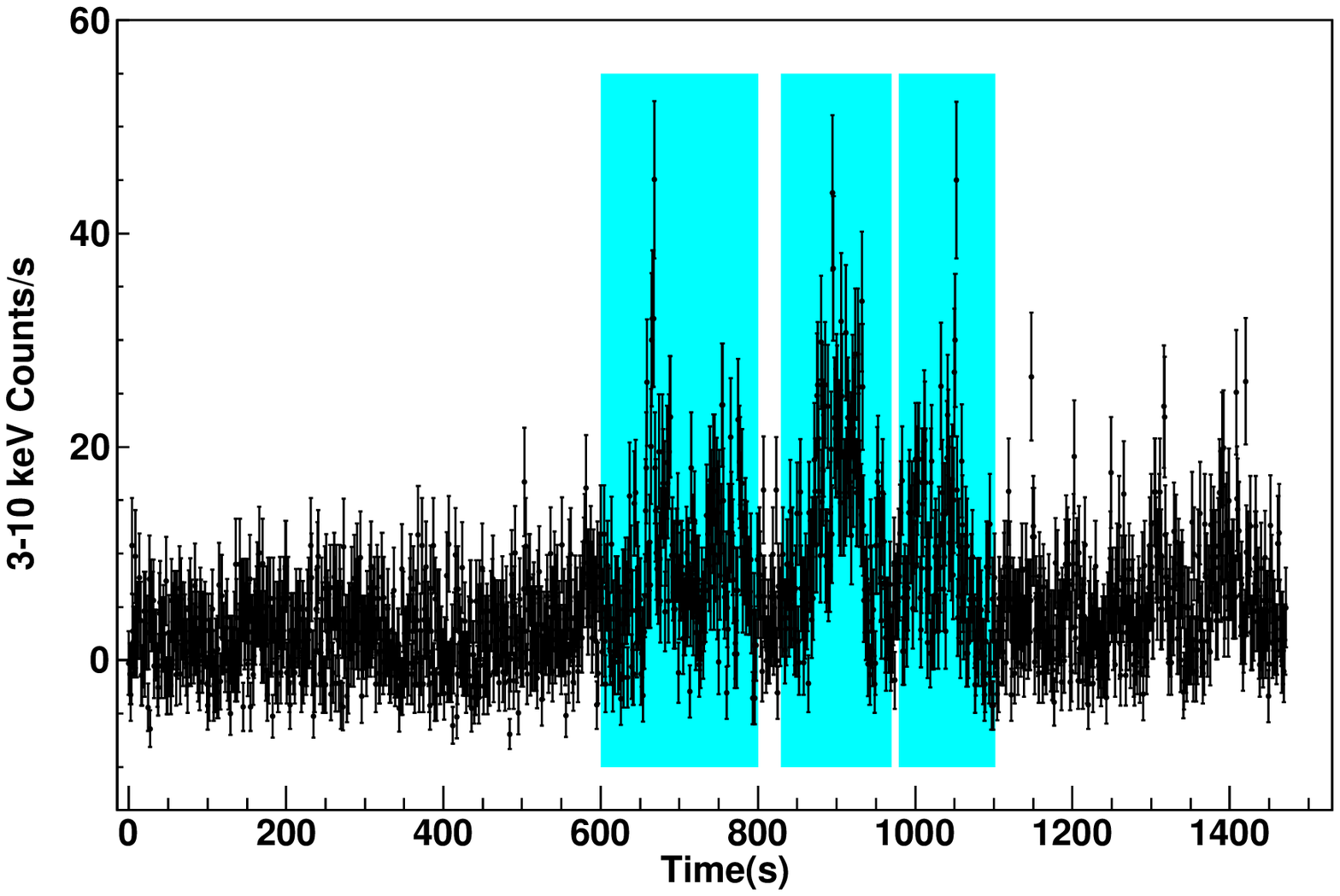}
    \includegraphics[angle=0, scale=0.38] {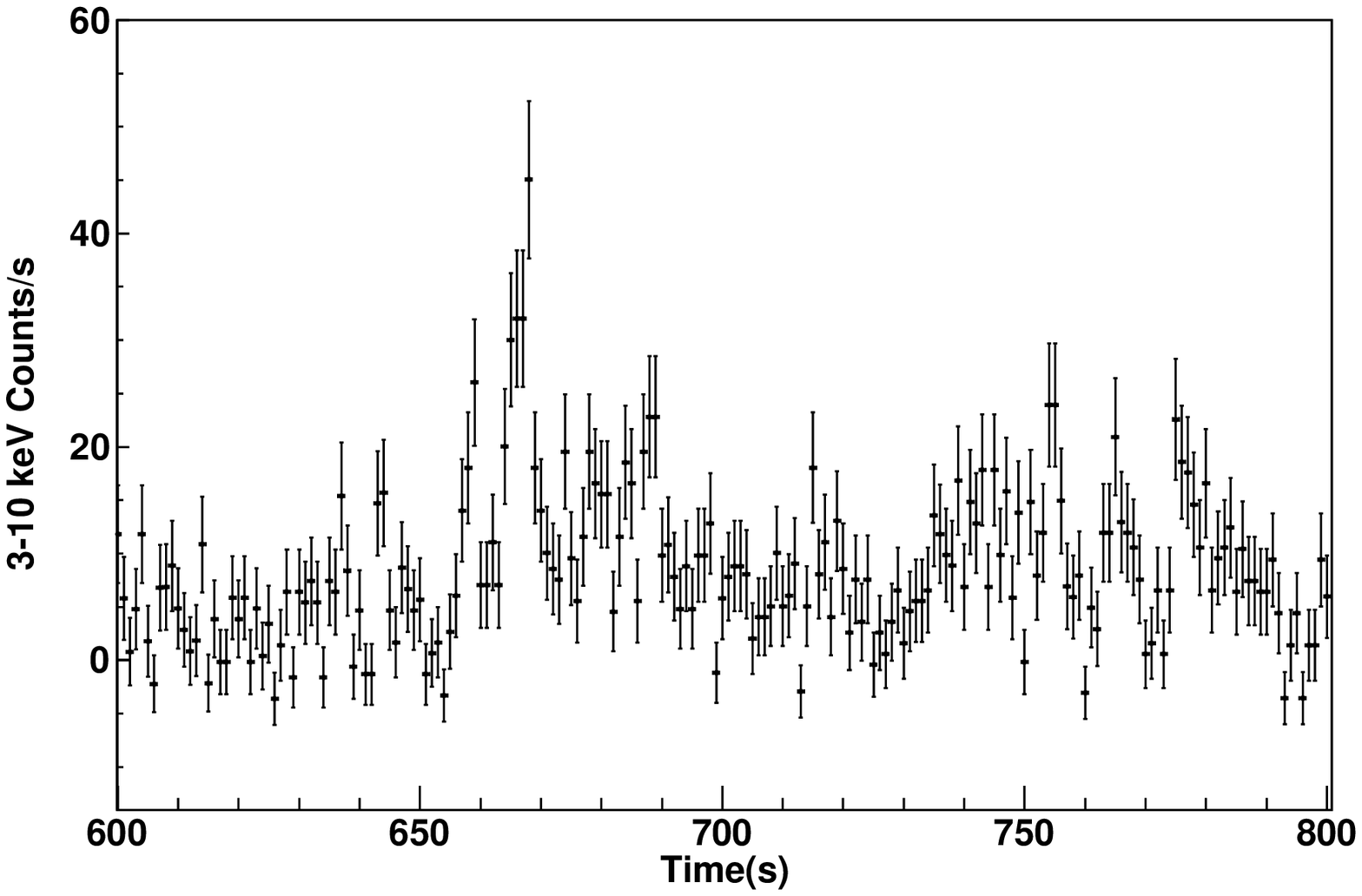}
    \includegraphics[angle=0, scale=0.38] {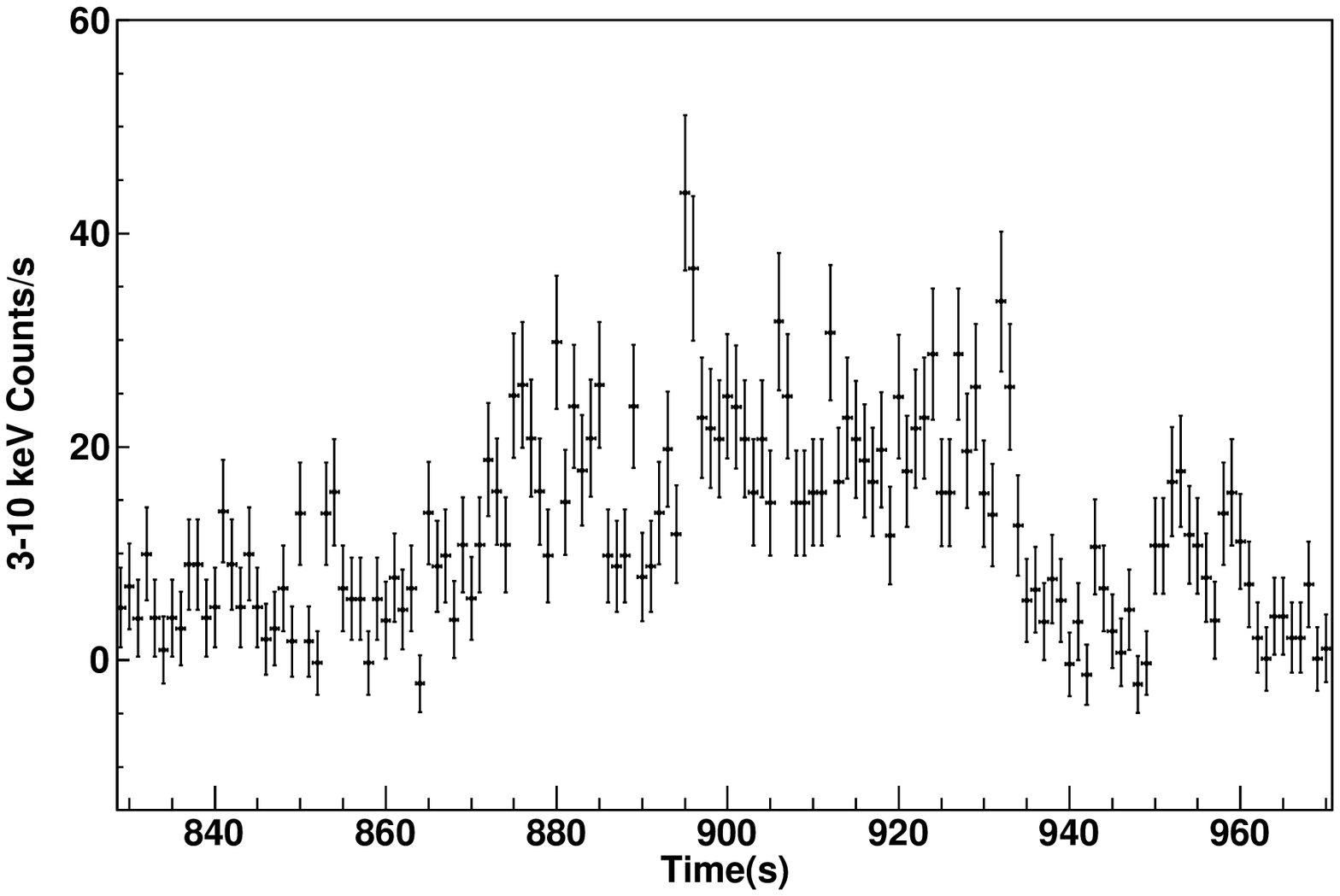}
  \includegraphics[angle=0, scale=0.38] {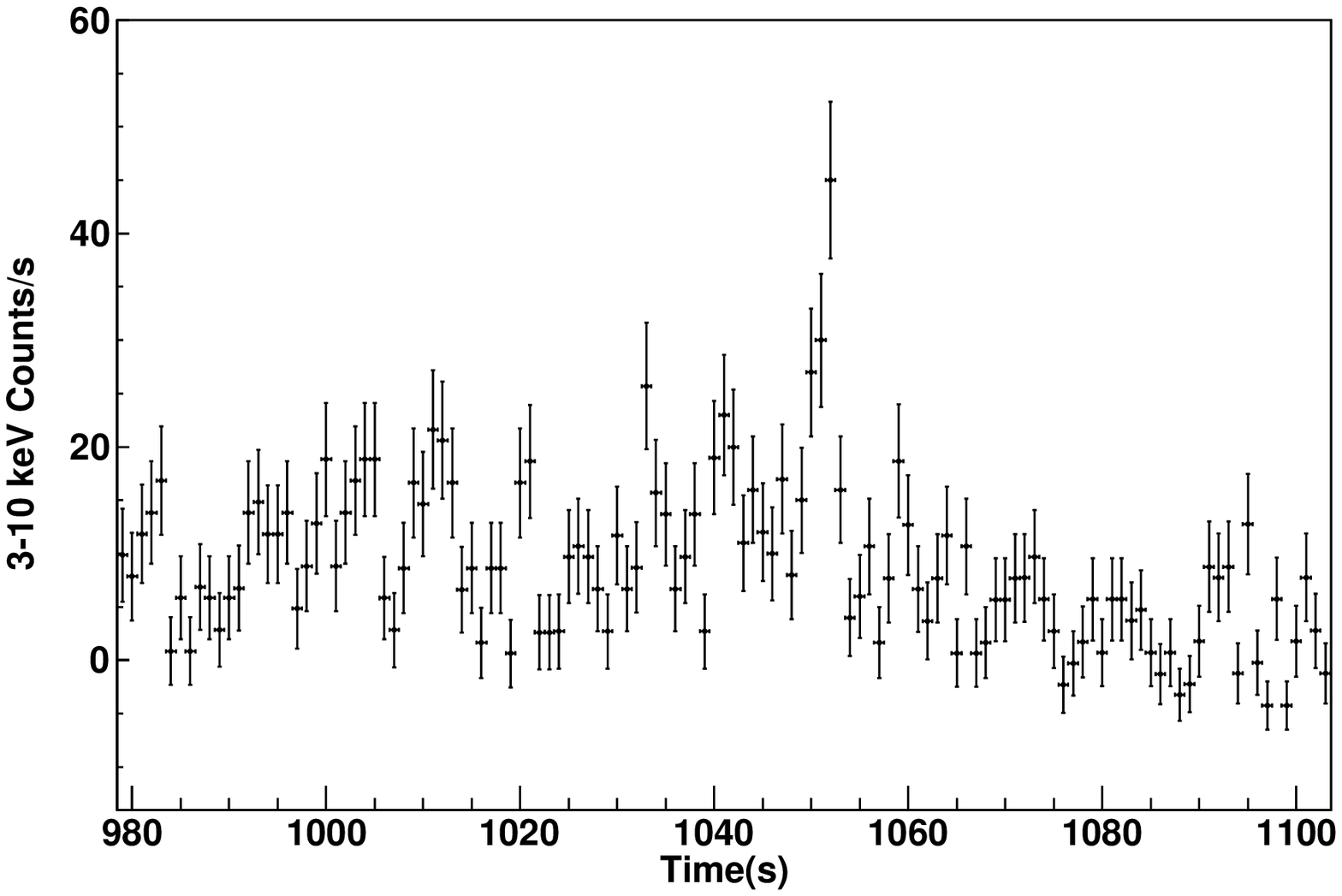}
     \caption{Top left panel: Lightcurve of the flare at MJD 54670.844444 (flare 4) binned in 1\,s
intervals with $t=0$ at the beginning of the time period; count rates are obtained from the top layer of PCU 2, 3 and 4 in the 3--10\,keV. The top right panel shows a zoom from $t=600$\,s to 800\,s. The bottom panels show zooms from 829\,s  to 970\,s and 979\,s to 1102\,s, respectively.
The color shadow in the top left shows the zoomed time spans. Note that the mean value of
the flux more than triples in 1 second from $t=894$\,s to 895\,s showing internal variability.}
     \label{f6}
\end{figure}

The inner structure of the flare at MJD 54670.844444 (flare 4) can be seen in Figure \ref{f6}, with a time bin of 1 s. It contains several sub-flares rising and falling in the timescale of 10--20\,s. The fastest flux variation lies between $t=894$ s and 895 s where the flux increases by more than 3 times in 1 second. This does not mean that the flare itself lasts for 1 s, rather, that it contains internal variability in a few seconds timescale. The total duration of the flare is similar to those previously found by other instruments, of about 1 ks, as discussed in the introduction and below. This can be seen easily in the top panel of Figure 1, also see the zoomed view presented in Figure \ref{flare-counts} where each point represent 64 s (top) or 16 s (bottom).

\begin{figure}[t]
\centering
    \includegraphics[angle=0, scale=0.4] {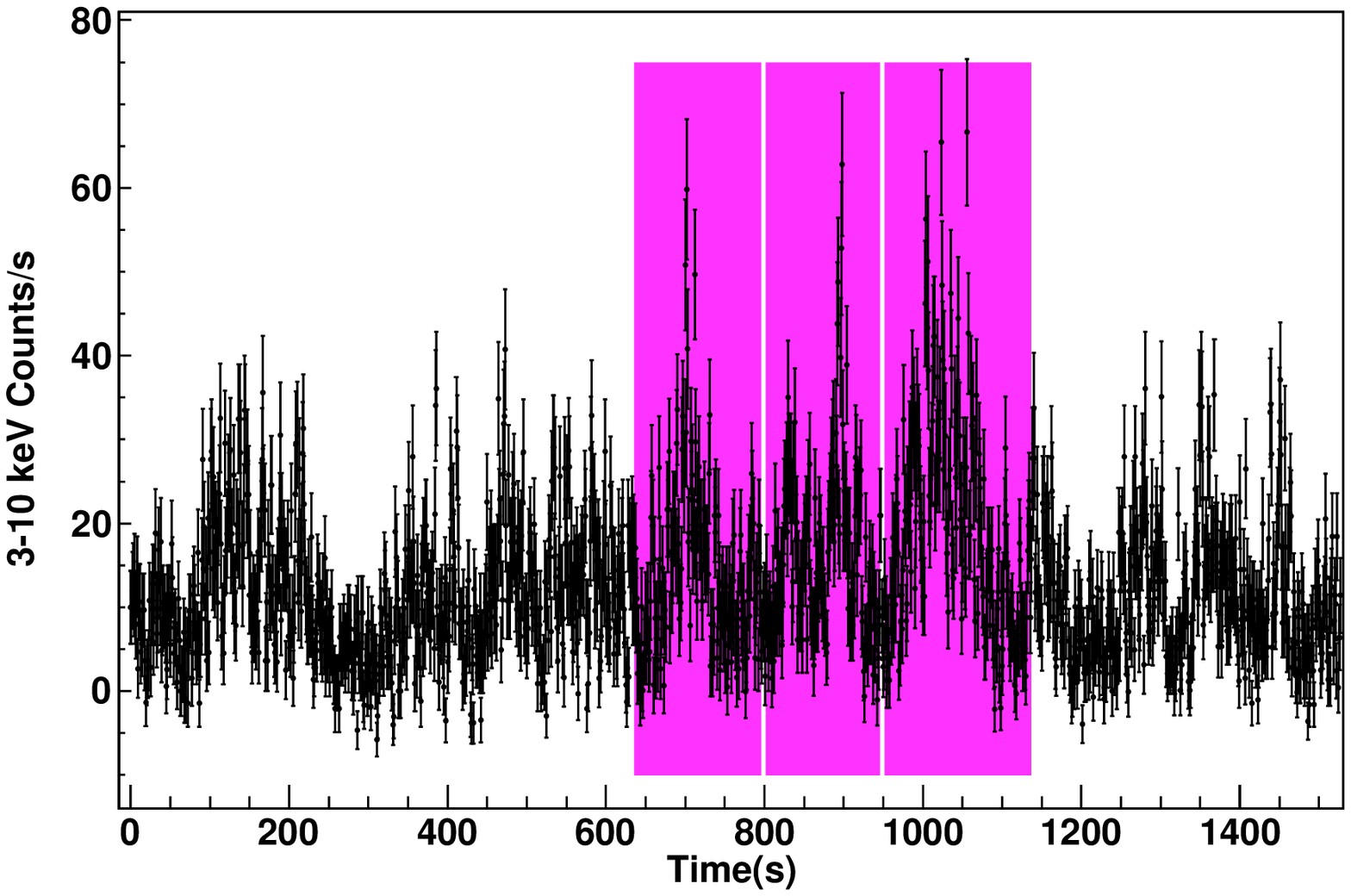}
  \includegraphics[angle=0, scale=0.4] {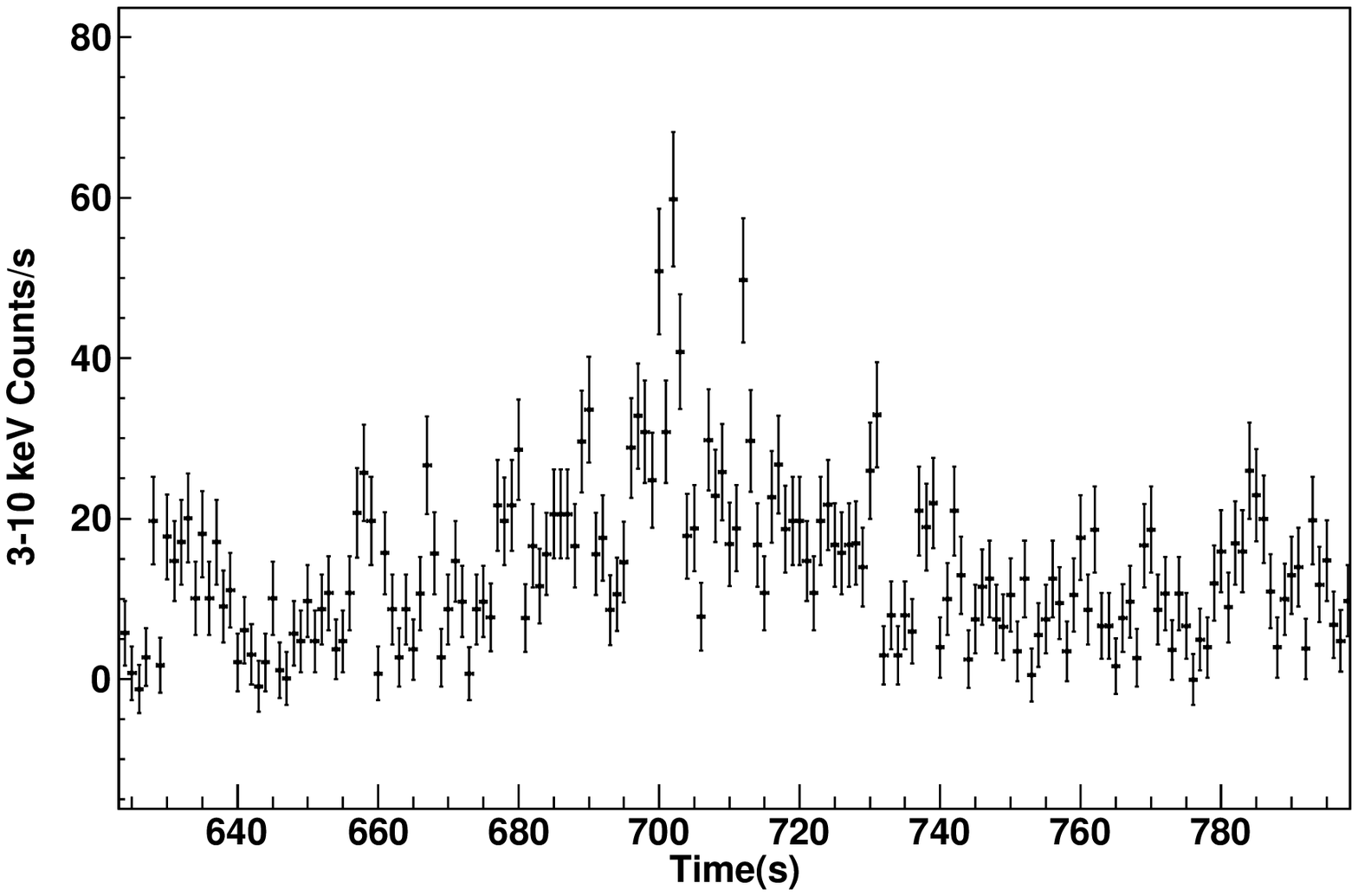}
    \includegraphics[angle=0, scale=0.4] {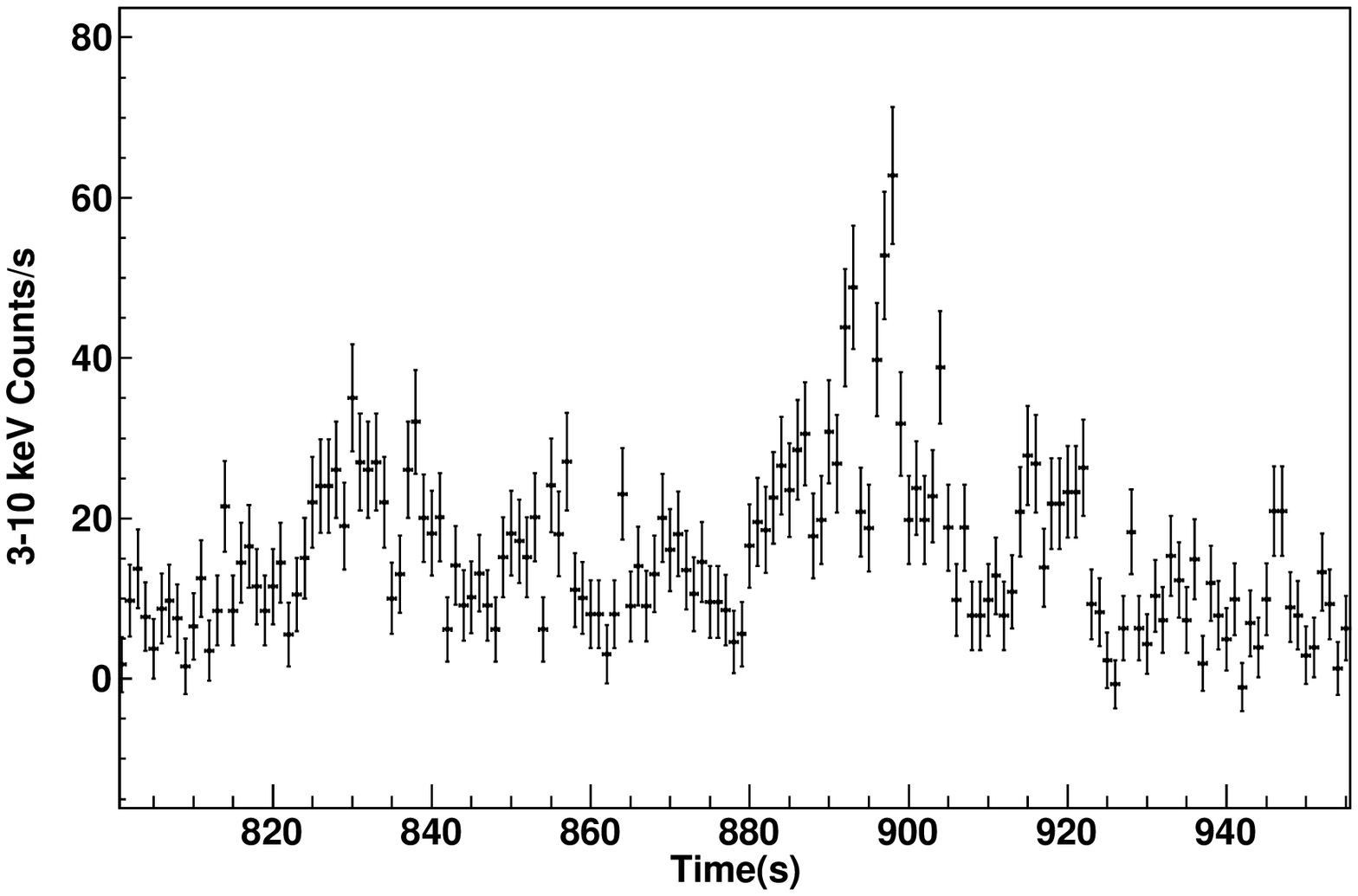}
      \includegraphics[angle=0, scale=0.4] {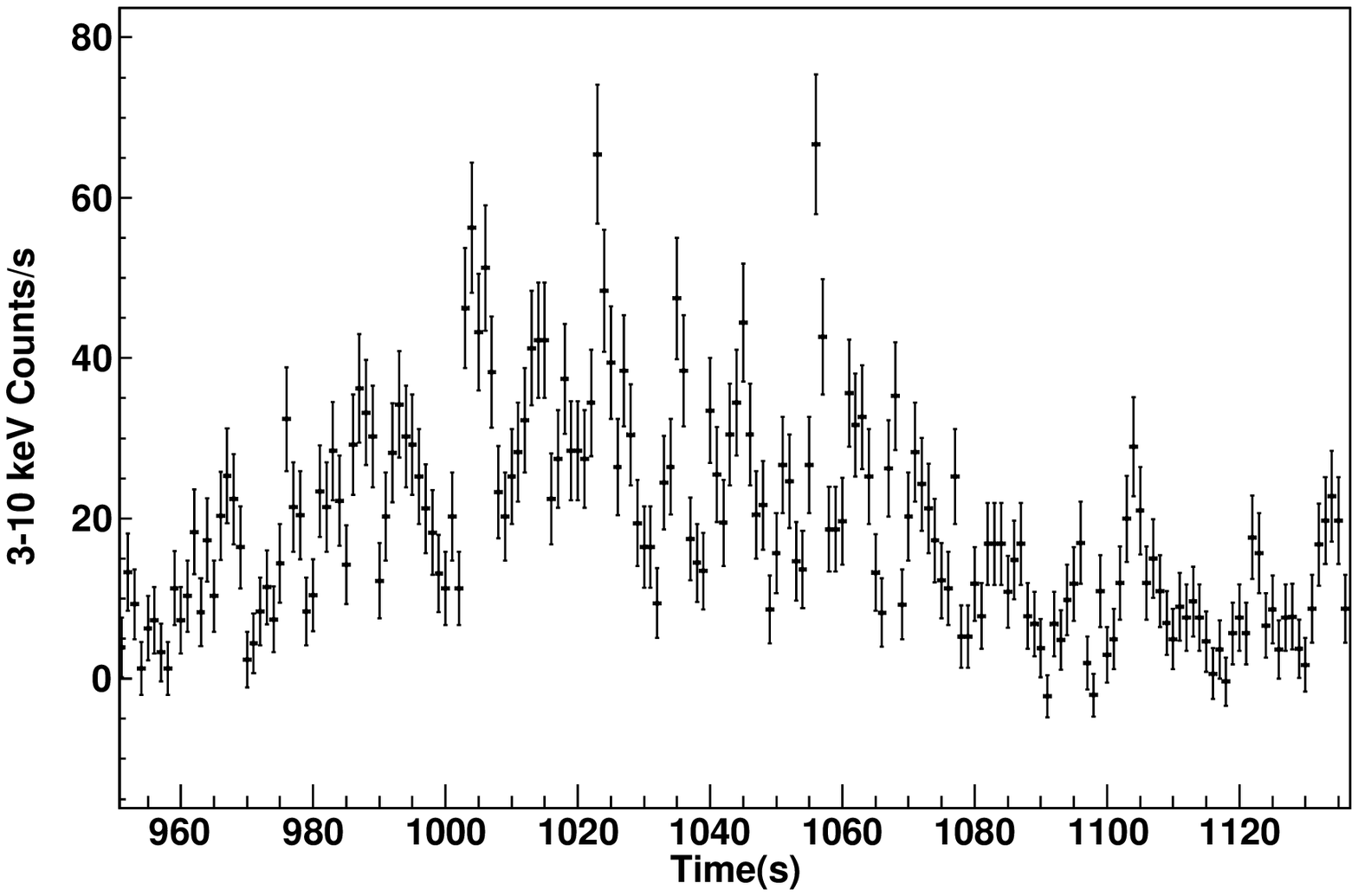}
     \caption{Top left panel: Lightcurve of the flare at MJD 54699.653333 (flare 5) binned in 1\,s intervals with $t=0$ at the beginning of the time period; count rates are obtained from the top layer of PCU 2, 3 and 4 in the 3--10\,keV. The top right panel shows a zoom from $t=600$\,s to 800\,s. The bottom panels show zooms from 800\,s  to 960\,s and 960\,s to 1120\,s, respectively. The color shadow in the top left shows the zoomed time spans, as in Figure \ref{f6}.
Note that
the flux more than triples in 1 second from $t=895$\,s to 898\,s showing internal variability.
}
        \label{f7}
\end{figure}

Similarly, the lightcurve of the flare at MJD 54699.653333 (flare 5), with 1 s time bins,  is shown in  Figure \ref{f7}. Also in this case, there is internal variability in the timescales of 10--20 s. The peak flux of this flare is higher than that appearing at MJD 54670.844444 (flare 4).
The strongest flux variation within flare 5 is situated between $t=$895 s and 898 s, where the flux increases more than four times in these three seconds. However, note that the size of the error bars in individual data points of 1 s are large.

\begin{figure}[t]
\centering
  \includegraphics[angle=270, scale=0.34] {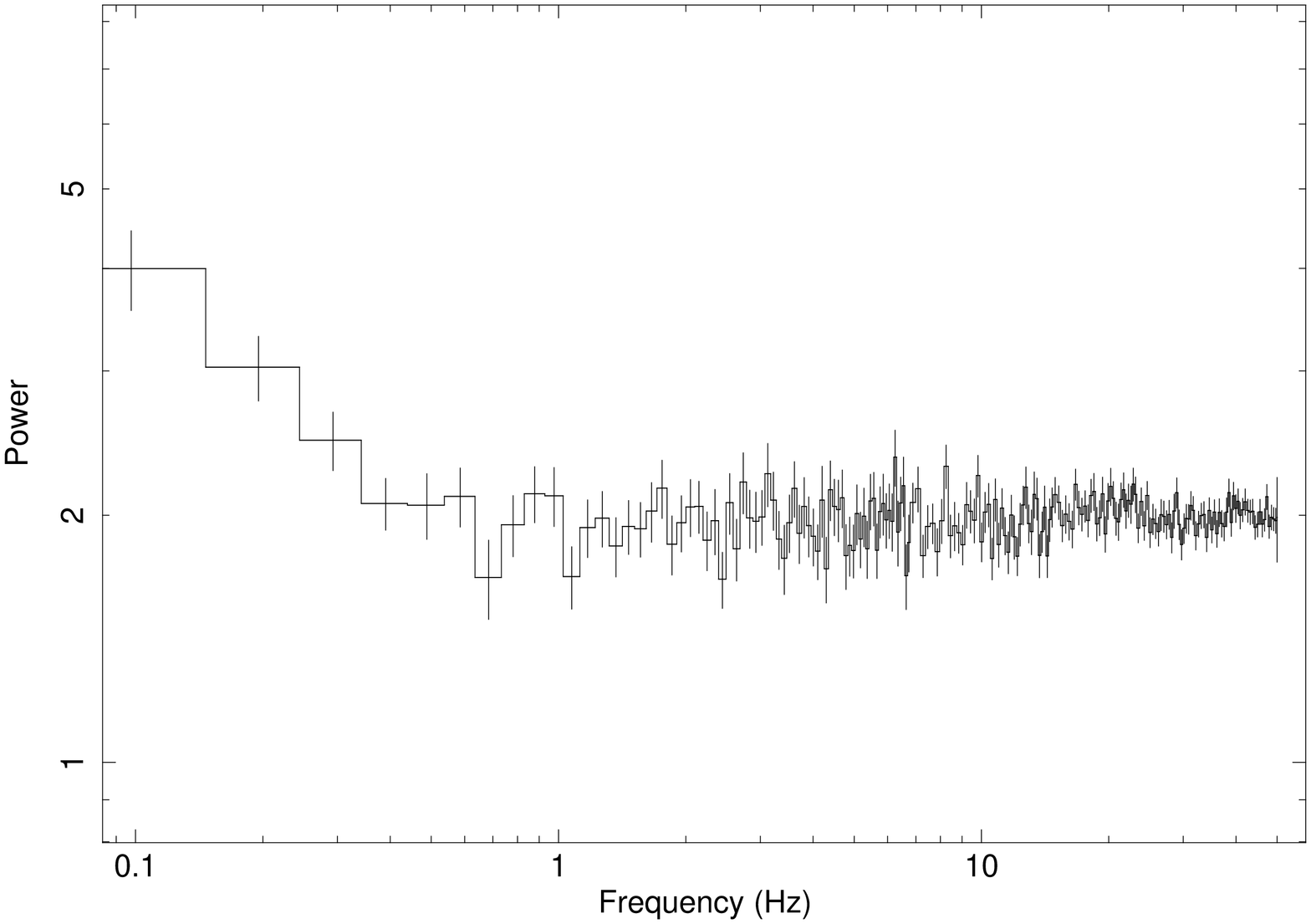}
  \includegraphics[angle=270, scale=0.34] {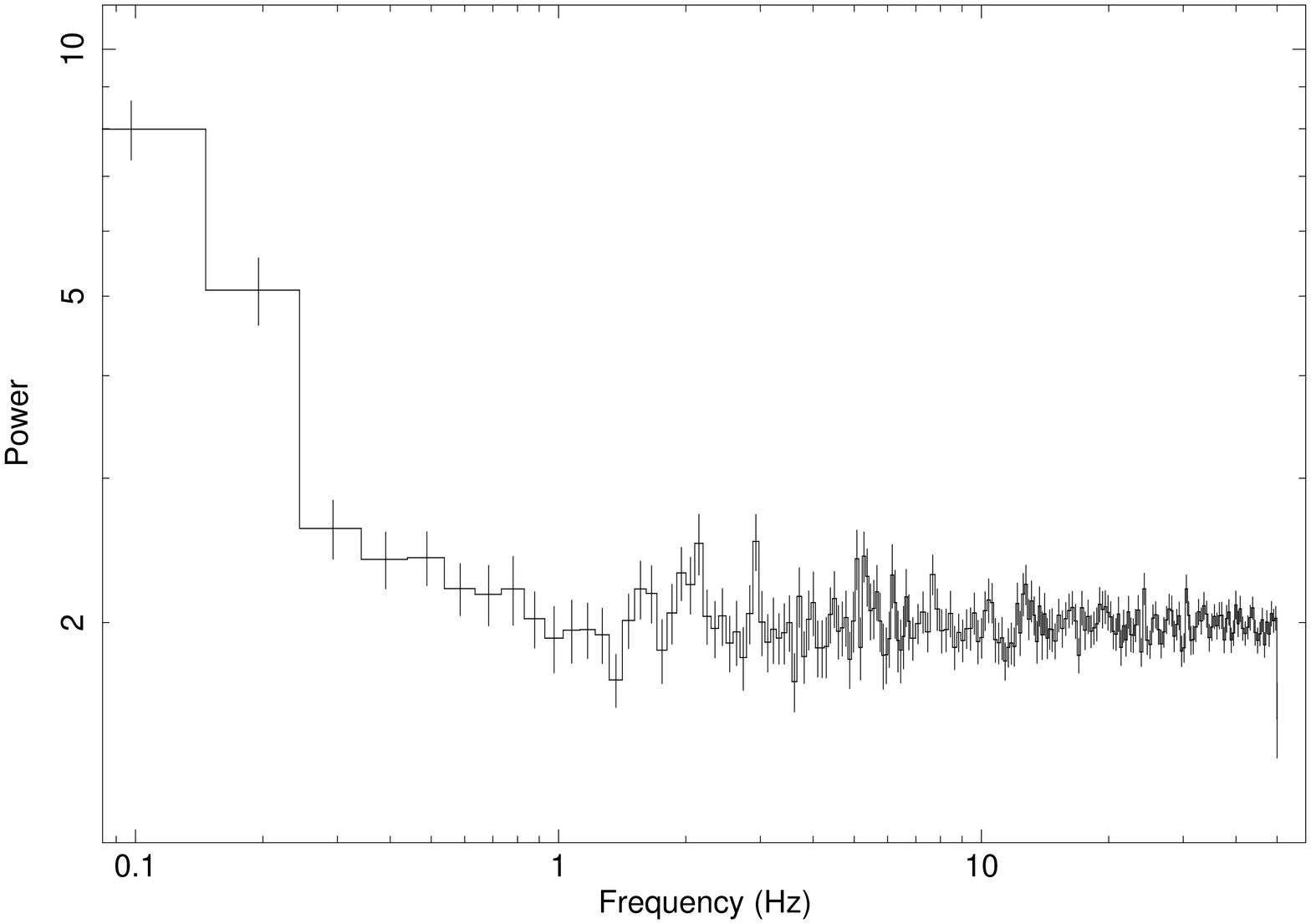}
     \caption{Left: Power spectrum of the flare found at MJD 54670.844444 (flare 4) as described in the text. Right: Power spectrum of the flare at MJD 54699.653333 (flare 5).
     No statistically significant structures are found. See the discussion in the text.}
\label{f8}
\end{figure}

To study the timing properties of both flares, we derive lightcurves from Good Xenon Data with time resolution at 0.01 seconds and then obtained a power spectrum. The result of timing analysis of these flares is shown in Figure \ref{f8}. The power spectrum for MJD 54670.844444 (flare 5) show no evidence for the existence of any structure in the power spectrum while the power spectrum for MJD 54699.653333 (flare 5) appears to show an QPO-like structure
at around 2 Hz.
In order to test the significance of this feature, we fit the power spectrum with two Lorenz profiles (one to match the low frequency rise) and a constant. The $\chi ^2$ of the fit is 177.4 under 189 degrees of freedom; with the  fitting parameters for the QPO being a
   central frequency  equal to $    2.115    \pm  0.015$, a
   width     of                    $ 0.103 \pm  0.259$, and a
   normalization    of       $ 0.698 $ (error cannot be well constrained). To test whether the QPO is significant we also fit the power spectrum with just one Lorenz profile (to match the low frequency rise) and a constant (i.e., removing the putative QPO), deriving  a $\chi ^2$ of 185.7 with 192 dof.
We carried out an F-test between the two fits.

The F-statistic value is  2.94 and the probability for rejecting the existence of the QPO is 0.034. Thus, we conclude that this seeming QPO is not significant.
Besides that, we explored whether the significance increases in different energy bands. The signal seems stronger in the 1.5--20 keV band, with and F-statistic value of 6.60 and rejecting probability at $2.897 \times 10^{-4}$, implying a pre-trail significance roughly of 3.5 $\sigma$. In this energy range other similar peaks do appear in the power spectrum, reducing the case for a real QPO appearing irrespective of the energy cut even further.
We caution that the existence of this QPO structure is mentioned in the ATel 1730 by Ray et al. (2008) and assumed as real in the
literature already, see Massi \& Zimmermann (2010). Our results discourage this assumption.

\subsection{Possible super-orbital influence}

The modulated flux fraction (obtained as ($c_{max} - c_{min} )/(c_{max} + c_{min}$ ), where $c_{max}$ and $c_{min}$ are the maximum and minimum counts rate in 3--30 kev in the orbital profile, after background subtraction) changes as a function of time. We here consider whether
it maybe correlated with the super-orbital radio period (1667$\pm$8 days, Gregory 2002). We take the reference time of the zero phase at MJD 43366.275 and the shape of the outburst peak flux modulation estimated by Gregory (2002) in radio. We fit this shape with a 1667$\pm$8 period sine function (see Figure 10, left panel), finding (not unexpectedly) that it could represent the radio super-orbital evolution well.

At each of the timescales considered (4, 5, 6, 7, and 9 months) we computed the modulated flux fraction from the counts rate in 3-30 keV as described above and found that as soon as timescales are long enough compared with the orbital period, so that orbit-to-orbit variations do not play a role, there is essentially a monotonically increasing trend of the modulated fraction. This is shown in the middle panel of Figure 10.

If we fit the modulation fraction values with the sine function mentioned above, no apparent correlation with the super-orbital radio period is found, at least with the data at hand (covering already $>0.5$ of the super-orbital period). For instance, the modulation fraction of each of the periods of 6-months were fitted with the above-mentioned sine function  (results can be seen in Figure 10, right panel, black line) and we derived a reduced $\chi^2$ of 14.04/4, much worse than reduced $\chi^2$ of just a straight line fit, 2.19/4.

\section{Discussion}

\subsection{On the spectral index versus flux}

The data collected with the PCA detector can be described (at all scales) by a featureless power-law spectrum with photon indices varying along the orbit, depending on the flux state.  Contrary to the typical low-hard and high-soft states of XRBs, both hosting a neutron star or a black hole, we confirm here earlier hints that there is a strong correlation between spectral index and flux presenting itself in the opposite direction. For \lsi, the spectral index value is anti-correlated with the flux level (otherwise stated, the spectral hardness is directly correlated with the flux level). The correlation seems linear, under scrutiny of current statistics, and present across all flux states of the source and all phenomenological timescales found. This anti-correlation was hinted at in {\it XMM} observations (48.7 ks, around orbital phase 0.6, Sidoli et al. 2006), and in {\it Chandra} (50 ks observations near orbital phase 0.04, Paredes et al. 2007, 95 ks observations from orbital phase 0.94 to 0.98); in addition of the already commented work by Smith et al. (2009) using PCA data (a sub-set of the study presented here). Earlier data is consistent with this now-proven trend; see, for example, the hardness ratio values plotted as figure 3 in the work of Taylor et al. (1996).
%
%

The spectral index - flux trend has also been found for another source, LS 5039, whose overall features across the electromagnetic spectrum make it very similar to \lsi.
Bosch-Ramon et al. (2005) found a similar tendency in fluxes and photon indices, although the data were background contaminated by diffuse X-ray emission from the Galactic Ridge (Valinia et al. 1998), likely preventing a proper photon index determination. Nonetheless, Takahashi et al. (2009) confirmed the trend, and showed a harder-when-brighter behavior of LS 5039 using Suzaku observations along its orbit. However, despite this similarity between \lsi\ and LS 5039, Kishishita et al. (2009) reported that the latter source has a X-ray lightcurve presenting long-term stability using multi-satellite observations between 1999 and 2007. For LS 5039, even fine structures in the lightcurves such as spikes and dips are found to be quite similar from one orbit to another, which is quite different from the \lsi\ behavior we find. The most appealing explanation for such a distinction is that the different behavior of the lightcurves arises from the stellar object, especially given the lack of knowledge about the nature of the compact object in both systems. While LS 5039 hosts an O star, \lsi\ has  a Be star primary, which is characterized with a dense equatorial disk. If, for instance, there are disc structure changes in time, the latter may induce variability in X-ray modulation. Admittedly, it could also be that the compact objects are different. Pulsar composed systems could provide feasible scenarios for a clocklike X-ray emission
(e.g., see Dubus 2006; Sierpowska-Bartosik \& Torres 2008), devoid of possible variability found in the low-hard state of black-hole composed systems where
the jet is thought to be connected to disc accretion, instabilities of which would lead to X-ray changes.

To address the issue of the origin of the X-ray emission, the fact that there is no accretion disc signature in our data, nor it is apparent in any other dataset of \lsi\ (see, for instance, Rea et al. 2010), as well as the lack of a cutoff in the X-ray spectra up to 30 keV as we find here, or in the {\it INTEGRAL} range (see Zhang et al. 2010), may not favour the X-ray radiation origin as thermal Comptonization of a corona. Therefore, our observations emphasize that the origin of the X-ray emission may be due to inverse Compton (IC) and/or synchrotron processes. The hardness increase with flux level may be explained differently depending on the underlying model for the system, but in principle both can accommodate it. In jet models like the one posed by Bosch-Ramon et al. (2006), an increase in accretion rate would imply a higher electron acceleration efficiency in the jets. There would be more particles at the higher-energy range of the electron energy distribution, and this would lead to a hardening of the X-ray spectrum. The fact that the maximum of the flux (the hardest spectrum) does not occur at periastron makes this interpretation more complex but still possible. On the other hand, in models where the X-ray emission arises as a result of IC scattering of ultrarelativistic electrons accelerated at the pulsar wind termination shock (for which losses scale with the distance to the massive star), it is the complex interplay of the different losses at hand with the acceleration efficiency that makes the maximal energy of electrons to vary along the orbit. Investigating the conditions for the particle propagation in the shock region requires setting the parameters defining the local magnetic field and the thermal field densities. The former scales with the distance from the pulsar site, whereas the latter scales with the distance from the massive star. Assuming \lsi\ parameters (even within the influence of their uncertainties, see, e.g., Sierpowska-Bartosik \& Torres 2009), and reasonable assumptions for the decay of the pulsar magnetic field out of the magnetosphere, one can compute all timescales for the losses involved (see e.g., Khangulyan et al. 2007 for a similar case in the system PSR B 1259-63); and use them to analyze the maximal energies of shock-accelerated electrons, finding that these are the highest for phases around apastron of \lsi. The extension of the electron spectra to higher energies may thus reflect on a harder, more luminous X-ray flux. In addition, if the pulsar wind moves through the region of varying density of the equatorial wind of the Be star, during its passage through the densest parts of the disk, X-ray emission may also be affected by Coulomb losses and adiabatic cooling of the accelerated particles, helping in the orbital modulation of the X-ray flux.

\subsection{On the flares}

Flares from \lsi\ at timescales of hours are known since long. For instance, Harrison et al. (2000) detected source variations of about 50\% on timescales of half an hour.
Sidoli et al. (2006) also found evidence for a rapid (again, on timescales of few hours) change in flux and hardness ratio: the source flux changed by a factor of 3  within 1 ks in their {\it XMM}-Newton observations, and by a factor of $\sim 4$ within 1.5 ks in their Beppo-SAX observations.
Esposito et al. (2007) found a similar behavior in Swift/XRT data, and
Paredes et al. (2007) found another miniflare over a timescale of  1 ks (with a total duration of 3 ks),
with the count rate increasing by a factor of 2. Rea et al. (2010) found, in {\it Chandra} data,  two flares lasting 3 and 1.5 ks with a rise-time of approximately 100 s. Due to their high angular resolution, {\it Chandra} and {\it XMM} observations provide certainty that the origin of this variability is \lsi.

We also observed significant flux fluctuations
in our PCA data. In total, 5 flares were detected by PCA and are shown in Figure 1 both in count rate and in fitted flux. Smith et al. (2009) independently reported on some of them, the earliest, while analyzing the initial part of this dataset. The overall features of all PCA flares are very similar one to another: in 1 s-binned lightcurves there is significant structure, we can see the flux rising up to 3 times in a single or a few bins (albeit each point has large error bars); with the whole flare lasting up to the order of ks.
Yet, the shortest of the bursts detected on \lsi\ observations has been made with
Swift-BAT, which reported the detection of a short (0.23 s), 10$^{37}$ erg (if at the distance of \lsi) in the 15--150 keV energy range (Barthelmy et al. 2008). Swift-XRT showed no increased flux from the direction of \lsi, just 921 s later. Was this an unlucky coincidence of a short unrelated GRB seen through the galaxy? Or was this emission also coming from \lsi? Was this magnetar activity (see Dubus \& Giebels 2008)? And in the latter case, how is the wind needed in all currently known pulsar models generating TeV emission maintained? Why is there radio emission from \lsi, unlike all known magnetars? Where is the region of generation of accelerated particles? Maintaining that \lsi\ is the first magnetar in a binary might necessarily lead to a complete reworking of the basic ideas for high-energy emission generation, since the pulsar wind zone or the pulsar wind -- stellar wind shock-induced particle acceleration cannot stand vis-a-vis.

Magnetars do show X-ray flaring activity on several time-scales, but they are not rotational-powered; their emission (flaring and persistent) is due to the presence and instability of their ultrastrong magnetic fields (see Mereghetti 2008 for a recent review).\footnote{It is interesting to note in passing that a similar $\sim$1 ks timescale variability has been found --also using PCA-- in the high mass X-ray binary 2S 0114+650 and it was also ascribed to a clumping wind (Farrel et al. 2008). 2S 0114+650 contains a super-slow rotator, one of the slowest spinning X-ray pulsars yet discovered (2.7 hours), which have also been assessed as a possible magnetar (Li \& van den Heuvel 1999).}  No magnetar has been found in GeV gamma-rays (Abdo et al. 2010); less in TeV. However, the models of (accreting) magnetars in binaries have been presented by Bozzo et al. (2008) and Bednarek (2009). In Bednarek (2009) models, the matter from the stellar wind penetrates the inner magnetosphere of the magnetar up to the point in which the magnetic pressure balance the gravitational pressure of the accreting matter creating a turbulent region, suggested to be prone to the acceleration of electrons to relativistic energies (the propeller phase). General features of the model seem in accord with the overall phenomenology, albeit the X-ray and the GeV emission are overpredicted in comparison with Fermi and our {\it RXTE} data.

In the accretion scenario flares are easily explainable as changes in the accretion rate from the companion star or the accretion disc about the compact object. Flares of such different timescales are not uncommon in accretion-based systems such as microquasars. In the rotational-powered pulsar scenario, the variability we observe would require clumpiness of the Be stellar wind, where the size, mass, and number density of the clumps imprint their signal on the X-ray emission (see Zdziarski, Neronov \& Chernyakova 2008, Rea et al. 2010). It does not appear contrived that clumps of the order of 10$^{11}$cm or smaller could be related with the variability found in PCA data, and similarly, with 1-ks timescales. For smaller timescales, other processes --likely depending on the nature of the companion-- probably have to be invoked.
We note that X-ray observations of LS 5039 (e.g., Bosch-Ramon et al. 2005) hourly timescale flux variations are similar to what we find for \lsi, but not faster variability.

\subsection{On the variability of orbital profiles}

As it shown here and in Paper I, orbital profile variability is seen from orbit-to-orbit all the way up to multi-year timescales. The phase of the profile maximum also changes.
The study of short-term, simultaneous multifrequency observations such as the one made by the MAGIC Collaboration (Anderhub et al. 2009) showed an X-ray -- TeV correlation. Is it sustained in time? In Paper I we have already noted that given the variability found in X-rays, the correlation shown by Anderhub et al. (2009) in about half an orbit produces local-in-time information useful for determining the process or the primaries responsible for the radiation detected, but probably it is not enough to establish an overall behavior. In fact, given the variable nature of the X-ray emission, a correlation found in a short observation might not be a sufficient proof that this correlation maintains with time. However,  if the TeV and X-ray emission are indeed always correlated, the TeV maximum should also vary in phase, affected by the different level of absorption at which the maximal photon production happens.
TeV and X-ray correlations could be explained, for instance, assuming dominant adiabatic losses (see, e.g., Zabalza et al. 2010). However, if such is the case, it would be highly unlikely to have whole orbits with only upper limits in the TeV range, while maintaining the X-ray emission level. This is apparently the case, as recent observations of VERITAS and MAGIC has shown that the source can turn off at TeV energies (Aliu et al. 2011, Jogler 2010). In fact, no TeV emission has been reported  since summer 2008, when the source re-appeared at periastron, emphasizing the variability of the TeV maximum phase-position (Ong et al. 2010). The adiabatic-losses one-zone model  has in addition the caveat of severely overproducing (unless a low-energy cutoff of the electron distribution is introduced ad-hoc) the GeV fluxes measured by Fermi (by 2-3 orders of magnitude). However, the 1--10 GeV data points by Fermi are at a much lower level than predicted by models, in general.

\begin{figure}[t]
\centering        
\includegraphics[angle=0, scale=0.271] {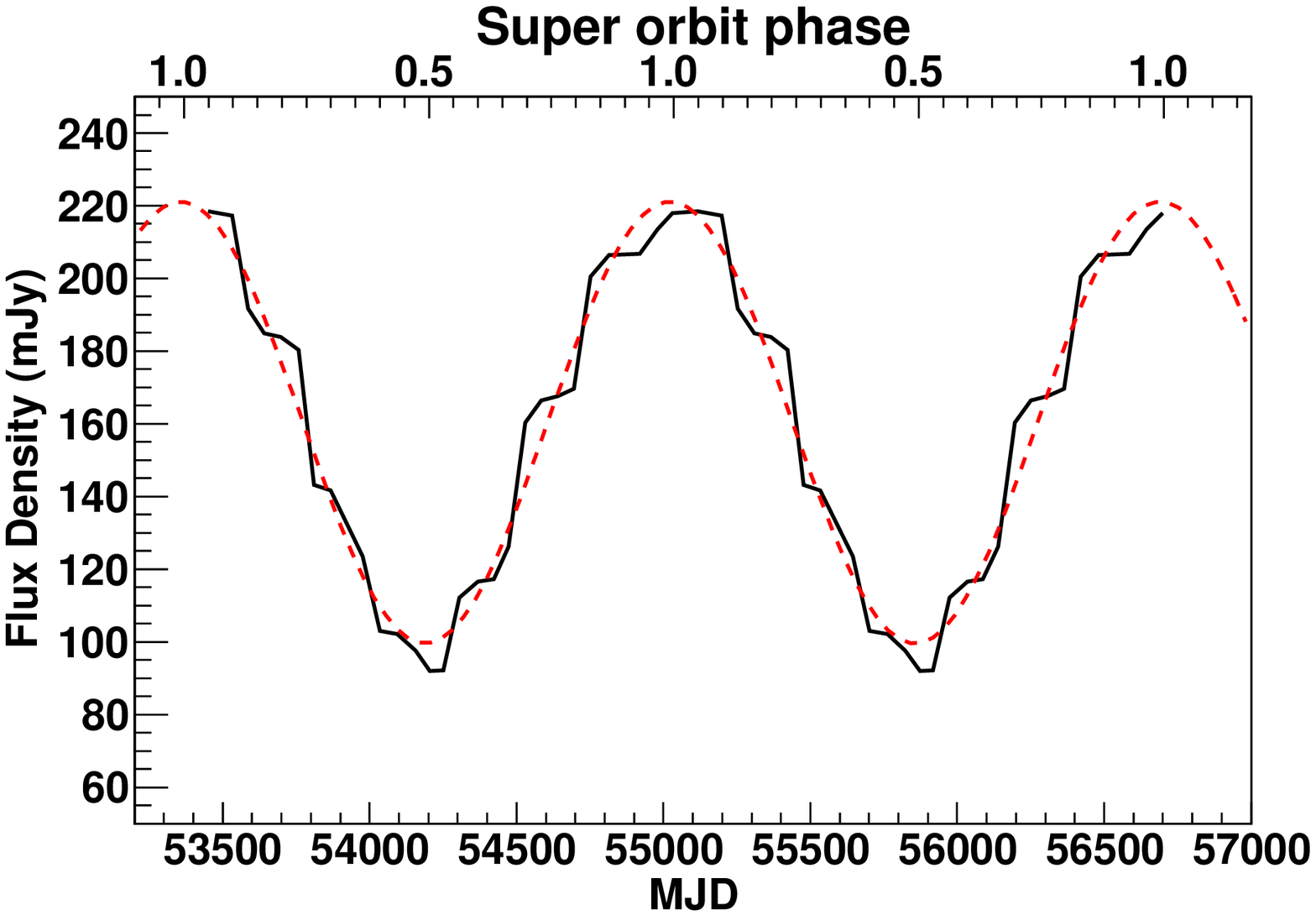}
    \includegraphics[angle=0, scale=0.271] {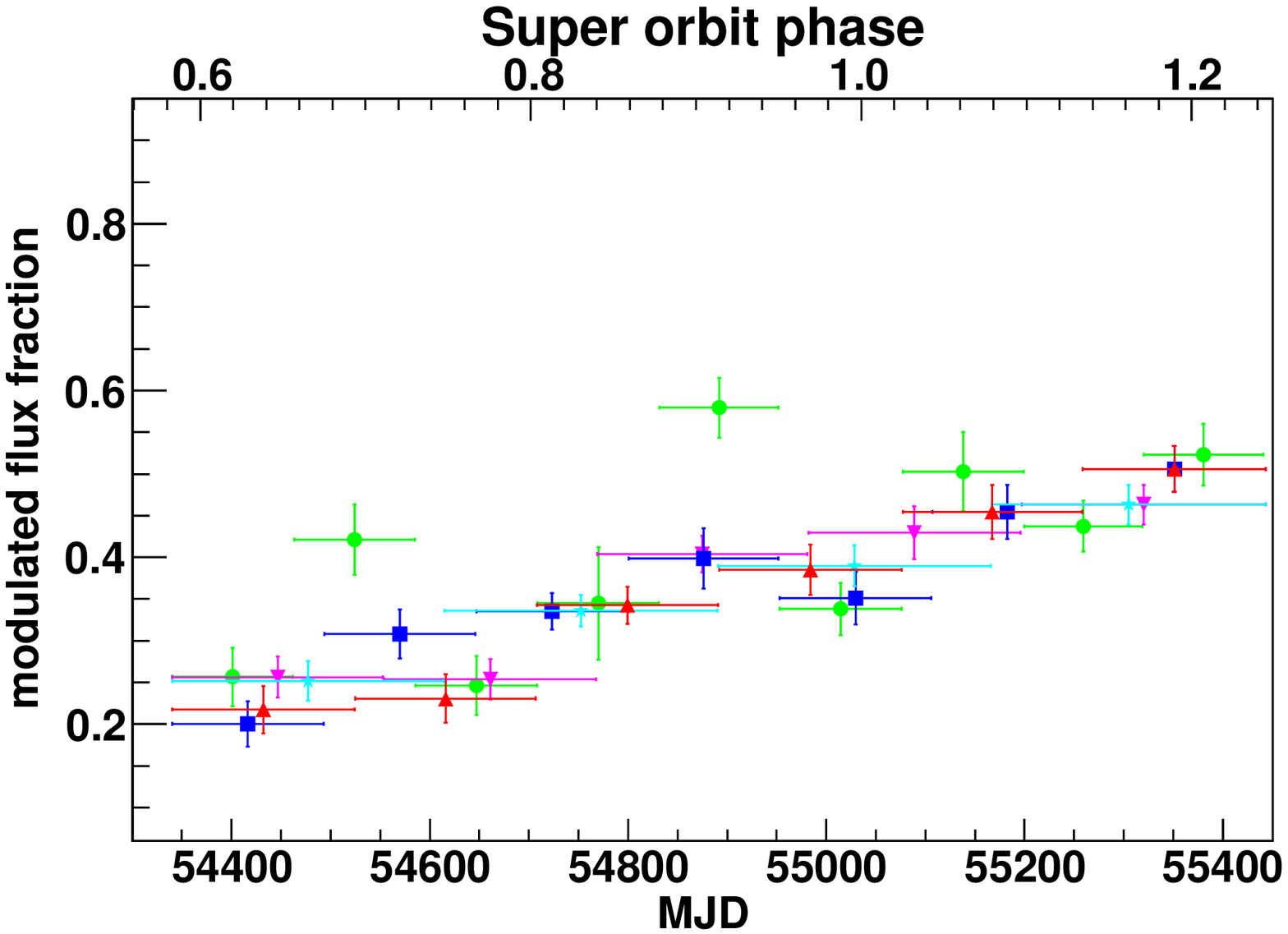}
        \includegraphics[angle=0, scale=0.271] {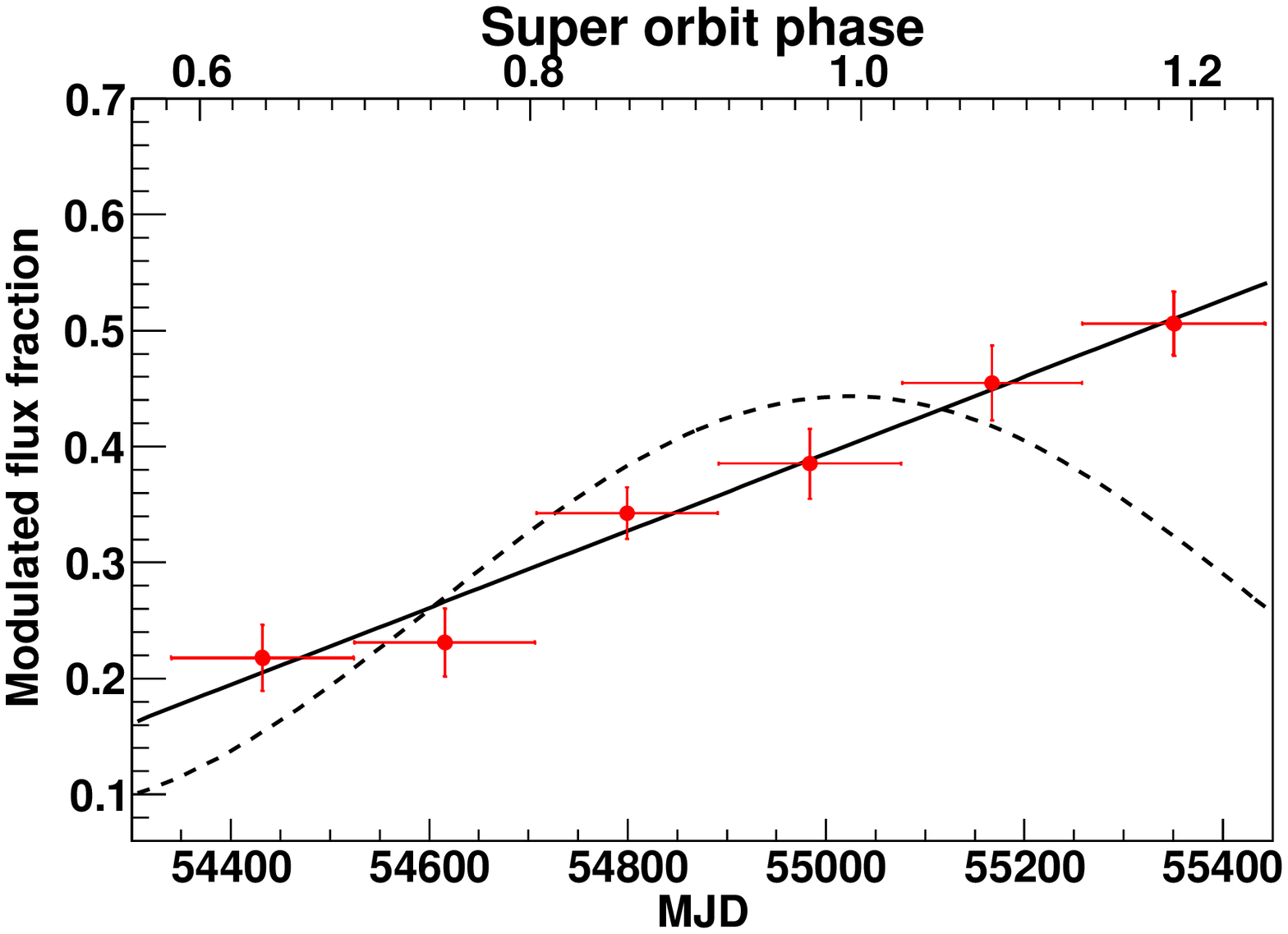}
         \caption{{The left panel shows the shape of the peak flux modulation in radio (black line) and the fitting  with a sine function (dotted red line). 
 The middle panel shows the evolution of the modulated fraction at different timescales (4, 5, 6, 7, and 9 months). The size of the {\it error} along the x-axis of each of the data points gives account of the timescales considered. The black line in right panel shows the linear fitting to the 6-months modulated flux fraction data. The dotted line in right panel shows the fitting with the sine function in left panel.}
}
\label{new-fig.}
\end{figure}

\section{Summary}

We reported here on the full analysis of a Rossi X-ray Timing Explorer ({\it RXTE}) Proportional Counting Array (PCA) monitoring of the $\gamma$-ray binary system \lsi. The data set covers 42 contiguous cycles of its orbital motion. A detailed discussion on the count rate variability across different timescales can be found in Paper I; although we have extended further this analysis with  additional orbits in order to consider the possible appearance of the super-orbital modulation. In here, we focused on the spectral properties and the analysis of two new flares appearing in these data set. The following is a summary of our findings.

\begin{itemize}

\item The soft X-ray emission from \lsi\ presents a periodic behavior at the orbital period, whose shape varies at all timescales explored. Profile variability is seen from orbit-to-orbit all the way up to multi-year timescales. The phase of the profile maximum also changes (see Paper I for an initial analysis in this sense). The study of simultaneous (part of one orbit) multifrequency observation showed a X-ray -- TeV correlation, but no evidence that it remains valid in time (Anderhub et al. 2009). TeV and X-ray correlations could be explained, for instance, assuming dominant adiabatic losses (see, e.g., Zabalza et al. 2010). However, if such is the case,  it  would be unlikely to have whole orbits with only upper limits in the TeV range, while simultaneously maintaining the X-ray emission level -- as found here to be the case along several years.

\item There were a total of 5 flares found on top of the typical behavior of \lsi. These flares, due to the large PCA field of view, may
or may not be associated with \lsi, but given earlier {\it XMM} and {\it Chandra} observations finding similar ks phenomenology, it is likely they are. 4 out of the 5 flares are grouped in the 0.6--0.9 orbital phase bin of \lsi. If the flares do not correlate with the orbital phase, the possibility of the resulting configuration is $2.76 \times 10^{-3}$. This may be indicative of an association, but certainly not a compelling proof. The inner structure of the flares present significant variability in shorter-than-1 ks timescales. Timing analysis of the flares does not show any structure in the power spectrum. The previously claimed QPO in the 5th flare is not statistically significant. Flares can be generated by changes in the mass loss rate of the star
or clumpiness of the Be stellar wind.

\item There is moderate variability of the spectral index across the 3 year observation period. We significantly confirmed the hints for an anti-correlation (now at $\sim10.2\sigma$) between spectral index and flux, detecting a harder-when-brighter behavior of the source. This could be explained differently by jet models and pulsar wind models. Our observation emphasizes that the X-ray emission may be due to inverse Compton and/or synchrotron processes, rather than thermal Comptonization of a corona.

\item {We explored the possible appearance of the super-orbital influence on the modulated fraction of the X-ray emission, with no clear detection of such a modulation in the X-ray band. Future data may perhaps clarify on this issue.}

\end{itemize}

\acknowledgements
This work was subsidized by the National
Natural Science Foundation of China, the CAS key Project KJCX2-YW-T03,
and 973 program 2009CB824800. Jianmin Wang and Shu Zhang
thank the Natural Science Foundation of China for support via
NSFC-10325313, 10521001, 10733010, 11073021 and 10821061. We
acknowledge support from the grants AYA2009-07391 and SGR2009-811, as
well as the Formosa Program TW2010005.
This work was also partially supported by NASA DPR No. NNG08E1671.
Nanda Rea is supported by a Ramony Cajal Fellowship.
We thank the referee for a careful reading of the manuscript.

\end{document}